\documentclass[12pt]{article}
 \pdfoutput=1

\usepackage{amsmath,amssymb,amsfonts}
\usepackage{hyperref}
\usepackage{graphicx}

\newcommand{\fref}[1]{Fig.~\ref{f.#1}} 
\newcommand{\eref}[1]{Eq.~(\ref{e.#1})}

\newcommand{\cref}[1]{Chapter~\ref{c.#1}}

\newcommand{\nn}{\nonumber \\}

\newcommand{\beq}{\begin{equation}} 
\newcommand{\eeq}{\end{equation}} 
\newcommand{\ba}{\begin{array}}  
\newcommand{\ea}{\end{array}} 
\newcommand{\bea}{\begin{eqnarray}}  
\newcommand{\eea}{\end{eqnarray} }  
\newcommand{\be}{\begin{eqnarray}}  
\newcommand{\ee}{\end{eqnarray} }  
\newcommand{\bal}{\begin{align}}
\newcommand{\eal}{\end{align}}   
\newcommand{\bi}{\begin{itemize}}  
\newcommand{\ei}{\end{itemize}}  
\newcommand{\ben}{\begin{enumerate}}  
\newcommand{\een}{\end{enumerate}}  
\newcommand{\bc}{\begin{center}}
\newcommand{\ec}{\end{center}} 
\newcommand{\bt}{\begin{table}}
\newcommand{\et}{\end{table}}  
\newcommand{\btb}{\begin{tabular}}
\newcommand{\etb}{\end{tabular}}  
\newcommand{\bvec}{\left ( \ba{c}}
\newcommand{\evec}{\ea \right )}

\def\cl{{\mathcal L}}  
   
\def\cm{{\mathcal M}}  
\newcommand{\cM}{{\mathcal M}}  
  
\def\co{{\mathcal O}}   
\def\cN{{\mathcal N}}

\def\tev{\, {\rm TeV}}

\def\pa{\partial}  
\newcommand{\re}{{\mathrm{Re}} \,}
\newcommand{\im}{{\mathrm{Im}} \,}
\newcommand{\tr}{\mathrm T \mathrm r}

\newcommand\simlt{\stackrel{<}{{}_\sim}}
\newcommand\simgt{\stackrel{>}{{}_\sim}}

\newcommand{\ti}{\tilde}  
\def\hc{{\rm h.c.}} 
\def\ov{\overline}  
  
\newcommand{\eps}{\epsilon}

\numberwithin{equation}{section}

\begin{document}
\begin{titlepage}
\vspace{-1cm}
\begin{flushright}
\small
CERN-PH-TH/2011-185,\ 
DESY 11-132,\
LPT-ORSAY 11-666
\end{flushright}
\vspace{0.2cm}
\begin{center}
{\huge \bf
If no Higgs then what? }
\vspace*{0.2cm}
\end{center}
\vskip0.2cm

\begin{center}
{\bf  A.~Falkowski$^{a}$, C.~Grojean$^{b,c}$, A.~Kami\'{n}ska$^d$, S.~Pokorski$^{b,d}$, A.~Weiler$^{b,e}$}

\end{center}
\vskip 8pt

\begin{center}
{\it $^{a}$ Laboratoire de Physique Th\'eorique d'Orsay, UMR8627--CNRS,\\ Universit\'e Paris--Sud, 91405 Orsay, France}\\
{\it $^{b}$ CERN Physics Department, Theory Division, CH-1211 Geneva 23, Switzerland}\\
{\it $^{c}$ Institut de Physique Th\'eorique, CEA/Saclay, F-91191 Gif-sur-Yvette C\'edex, France}\\
{\it $^{d}$ Institute of Theoretical Physics, Faculty of Physics, University of Warsaw,\\ Ho\.za 69, 00-681, Warsaw, Poland}\\
{\it $^{e}$ DESY, Notkestrasse 85, D-22607 Hamburg, Germany}
\end{center}

\begin{center}
{\tt adam.falkowski@th.u-psud.fr, christophe.grojean@cern.ch, anna.kaminska@fuw.edu.pl, stefan.pokorski@fuw.edu.pl, andreas.weiler@desy.de}
\end{center}

\vspace*{0.3cm}

\vglue 0.3truecm

\begin{abstract}
\vskip 3pt \noindent

In the absence of a Higgs boson, the perturbative description of the Standard Model ceases to make sense above a TeV.
Heavy spin-1 fields coupled to W and Z bosons can extend the validity of the theory up to higher scales. 
We carefully identify regions of parameter space where a minimal addition 
---  a single spin-1 $SU(2)_{\rm custodial}$-triplet resonance --- allows one to retain perturbative control in all channels. 
Elastic scattering of longitudinal W and Z bosons alone seems to permit a very large cut-off beyond the Naive Dimensional Analysis expectation.
We find however that including scattering of the spin-1 resonances then leads to  an earlier onset of strong coupling.
Most importantly for LHC searches, we define a self-consistent set-up with a well-defined range of validity  without recourse to unitarization
schemes whose physical meaning is obscure. 
We discuss the LHC phenomenology and the discovery reach for these electroweak  resonances and mention the possibility of a nightmare scenario with no Higgs nor resonance within the LHC reach.
Finally, we discuss the effects of parity breaking in the heavy resonance sector which reduces the contributions to the $S$ parameter. 

\end{abstract}

\end{titlepage}

\newpage
\tableofcontents

\section{Introduction}\label{sec:intro}

The main goal of the LHC is to understand the dynamics of electroweak symmetry breaking and to discover the infamous Higgs boson. The LHC has been designed not to miss it if it exists and corresponds to its Standard Model incarnation. Its mass is indeed subject to various theoretical constraints such as the vacuum stability, the triviality and the perturbative unitarity bounds~\cite{Djouadi:2005gi} that, if it has maliciously escaped at the LEP~\cite{LEP} and the Tevatron~\cite{Tevatron}, guarantee its discovery at the LHC~\cite{cmstdr, atlastdr}. Furthermore the electroweak precision data preciously collected over the years require a delicate screening of the radiative corrections to the gauge boson propagators that can be accounted for only with a relatively light Higgs boson~\cite{LEPEWWG}. Of course Nature does not have to follow the minimal path envisioned by theorists and the conclusion that the LHC will for sure see something has to be reassessed in all possible alternatives. For instance if the Higgs boson is not an elementary particle but rather a composite bound state emerging from a strongly interacting theory, its discovery might require more patience/luminosity~\cite{Giudice:2007fh, Espinosa:2010vn}.

In this paper we discuss the prospects to observe the degrees of freedom that unitarize the $V_L V_L$ scattering amplitudes, $V=W,Z$, in the context of strong electroweak symmetry breaking saturated by vector resonances. Even though such scenarios are generically challenged by electroweak data,  they rely after all on some sort of dynamics that we know is realized in Nature both in condensed matter and in high energy physics.  Our study will be guided  by a symmetry principle, an approximate $SU(2)_C$ custodial symmetry to avoid undesirably large deviations to $\rho=1$, and by a dynamical assumption inspired by QCD, namely vector meson dominance~\cite{Ecker:1988te}, ie the saturation of the amplitudes by the lightest vector resonances rather than by other types of resonances or by structureless dynamics. For spin-1 resonances, only a $SU(2)_C$ triplet can contribute to the $W_L W_L$ scattering amplitudes without inducing an excessive contribution to the $T$ parameter (see for instance Ref.~\cite{Alboteanu:2008my}). It was shown in Refs.~\cite{Csaki:2003dt, SekharChivukula:2001hz} that a tower of spin-1 resonances can postpone the perturbative breakdown of the $V_L V_L$ amplitudes, provided that their masses and couplings satisfy certain sum-rules. We shall consider here a minimal setup with a single $SU(2)_C$ triplet resonance, $\rho$, in the electroweak symmetry breaking sector. Our main concern is what the perturbative unitarity requirement has to say about this minimal setup: What is the allowed mass of the resonance? What are the prospects to observe such a resonance at the LHC? What are its couplings to the light SM degrees of freedom? Up to which energy is the setup self-consistent? At which scale do we expect to see another resonance? Ie, what is the high energy behavior of the $V_LV_L$ amplitudes which are known to enter a non-perturbative regime between 1.2 and 3~TeV in the absence of any UV moderator~\cite{Lee:1977eg}.
  
The Higgs couplings in the Standard Model are such that unitarity is ensured in both elastic and inelastic channels~\cite{Cornwall:1974km, Contino:2010mh} up to arbitrarily high scale. This is not possible with spin-1 resonances and we shall see that important constraints are obtained from the inelastic channels $V_L V_L \to \rho \rho$ and $V_L \, \rho \to V_L\, \rho$ processes (see Ref.~\cite{Papucci:2004ip} for a discussion on unitarity in inelastic channels for a Higgsless model). While it is possible to delay the perturbative unitarity breakdown in $V_LV_L$ scattering by appropriately tuning the coupling to a $\rho$-resonance, the constraints on the inelastic channels prevent a perturbative description above the NDA cutoff of the SM without a Higgs boson, at least in the minimal setup with a single resonance triplet.

One legitimate concern about models of strong electroweak symmetry breaking is their consistency with electroweak precision data as well as with flavor constraints. Actually part of the trouble originates from the absence of the light Higgs contribution to the oblique parameter, which would then call for either a positive contribution to $T$ or a negative contribution to $S$. Unfortunately, the resonances contributing to $V_LV_L$ scattering tend to give an opposite sign contribution~\cite{Agashe:2007mc} and additional dynamics like degenerate axial vectors~\cite{Barbieri:2008cc} or composite fermions~\cite{Cacciapaglia:2004rb, Chivukula:2005bn, Barbieri:2008zt} is called on rescue.  Since the focus of this paper is the behavior of the $V_LV_L$ amplitudes, we shall not pay attention to these additional degrees of freedom here.

Even if the excesses recently reported by ATLAS and CMS~\cite{EPS2011} are the first signal of a light Higgs boson, our approach to the search for resonances in the $WW$ scattering, properly generalized, will be useful to distinguish  an elementary from  a composite scalar. And in the absence of any other signal of new physics, the measurement of the $WW$ scattering amplitude will be the only handle to decipher the true dynamics of the electroweak symmetry breaking.

\section{Electroweak Chiral Lagrangian interacting with $\rho$ mesons}\label{sec:lagrangian}

In this section we describe the interactions of the SM electroweak sector with the electroweak breaking sector.
The latter is assumed to have a low-energy effective description where the only degrees of freedom are:
\bi 
\item   3 Goldstone bosons $\pi$ who become the longitudinal polarizations of the $W$ and $Z$  bosons,  
\item    A triplet of massive vector bosons referred to as the $\rho$ mesons.  
\ei 
We assume the effective lagrangian for the electroweak breaking sector obeys an $SU(2)_L \times SU(2)_R$ global symmetry which is spontaneously broken to its diagonal subgroup $SU(2)_V$  and whose $SU(2)_L \times U(1)_Y$ subgroup is weakly gauged by the SM gauge bosons. 
The Goldstone bosons are  described by the non-linear sigma model field $U = e^{i \sigma \cdot \pi(x) /v}$ transforming as $U \to g_L U g_R^\dagger$ under the global symmetry.   
The couplings to $\rho$-mesons can be introduced in several ways. 
Here we follow Ref.~\cite{Bando:1984ej,Casalbuoni:1985kq}  where  $\rho$ is  the  gauge boson of a local ``hidden" $SU(2)_h$ symmetry.  
To this end one writes $U = \xi_L \xi_R^\dagger$ and assigns the transformation law $\xi_{L,R} \to g_{L,R} \xi_{L,R} h^\dagger$.   
The vector bosons can now be introduced via the covariant derivatives,   
\bea
D_\mu \xi_L &=& \pa_\mu \xi_L - i  {g \over 2}  W_\mu^a \sigma^a  \xi_L  + i {g_\rho  \over 2} \xi_L \rho_\mu^a \sigma^a   
\nn
D_\mu \xi_R &=& \pa_\mu \xi_R - i  {g' \over 2}  B_\mu  \sigma^3 \xi_R  + i  {g_\rho \over 2}  \xi_R \rho_\mu^a \sigma^a      
\eea 
where $g$,$g'$, $g_\rho$ are the gauge couplings of  $SU(2)_L \times U(1)_Y \times SU(2)_h$. 
We shall assume the strong sector coupling dominates, $g_\rho \gg  g$. 
One can define
\bea
V_\mu^{\pm} &=&   \xi_L^\dagger  D_\mu \xi_L \pm \xi_R^\dagger D_\mu \xi_R 
\eea  
that transform in the adjoint of  $SU(2)_h$, $V_\mu^{\pm} \to h V_\mu^{\pm}h^\dagger$. 
Under the parity symmetry exchanging $L \leftrightarrow R$,  once $\rho$ is assigned positive parity,  $V_\mu^+$ is even while  $V_\mu^-$ is odd.   
Assuming the electroweak breaking sector conserves parity, at the leading order in the derivative expansion only two terms are allowed in the lagrangian,    
\beq
\label{e.LeadingLagrangianPC}
 - {v^2 \over 4}  \tr \left \{   \alpha V_\mu^+ V_\mu^+  +  V_\mu^- V_\mu^-   \right \},  
\eeq 
Eq.~(\ref{e.LeadingLagrangianPC}) gives rise to gauge boson mass terms as well as kinetic and interaction terms involving Goldstone bosons. 
For $g_\rho \gg g$ the eigenvalues of the gauge boson mass matrix are hierarchical.
The largest eigenvalues $m_\rho \approx \sqrt{\alpha} g_\rho v$ set the mass scale of the $\rho$-meson triplet. 
The positivity of mass and kinetic terms implies that the parameter $\alpha$ must be positive but otherwise it is unconstrained. 
The lower  eigenvalues are  $m_W \approx g v/2$ in the charged sector and  $m_\gamma = 0$, $m_Z \approx \sqrt{g^2 + g'{}^2} v/2$ in the neutral sector. 
These are identified with the SM gauge boson masses, which  fixes the overall scale in \eref{LeadingLagrangianPC} to be $v  = 246$~GeV. The entire procedure of identifying physical degrees of freedom of the Lagrangian \eref{LeadingLagrangianPC} is described in detail in Appendix~\ref{sec:eigenstates}.
The kinetic terms for the gauge fields can be introduced at the $p^4$ level\footnote{The counting is $[\pa_\mu] = [A_\mu] = \co(p).$ }, 
\beq
- {1 \over 4} L_{\mu\nu}^a  L_{\mu\nu}^a  - {1 \over 4} B_{\mu\nu}  B_{\mu\nu}  - {1 \over 4 } \rho_{\mu\nu}^a  \rho_{\mu\nu}^a  
\qquad A_{\mu\nu}^a = \pa_\mu A_\nu^a - \pa_\nu A_\mu^a - g_A \eps^{abc} A_\mu^b A_\nu^c  
\eeq 
 
The Goldstone bosons can be conveniently parametrized as  
\beq 
\label{e.xiparam}
\xi_L = e^{\phantom{-}i \pi^a \sigma^a/2v}  e^{-i  G^a \sigma^a/2 v  \sqrt \alpha  } 
\qquad 
\xi_R = e^{-i \pi^a \sigma^a/2v}  e^{- i G^a \sigma^a/2v  \sqrt \alpha }  
\eeq
Here $\pi^a$ and $G^a$ are triplets of Goldstone bosons that become the longitudinal polarizations of $W$, $Z$  and  the triplet of $\rho$ mesons.
Generally speaking, one can define Goldstone bosons ``eigenstates" as the linear combinations of $\pi^a$ and $G^a$ that have diagonal kinetic terms and diagonal kinetic mixing with the gauge boson mass eigenstates. 
The parametrization in \eref{xiparam} is such that $\pi^a$ mixes only with the SM gauge bosons while $G^a$ mixes only with $\rho$, up to small corrections  suppressed by $g^2/g_\rho^2$ (see Appendix~\ref{sec:eigenstates}).
In the following we work in the unitary gauge for $\rho$ and set $G^a=0$ but we shall keep $\pi^a$. 
By the Goldstone boson equivalence theorem,  the scattering amplitudes of $\pi^a$, for $s \gg m_W^2$, are equal to the scattering amplitudes of longitudinally polarized $W$ and $Z$  bosons.   
The relevant interaction terms for computing these amplitudes are 
\beq
g_{\rho \pi \pi}  \eps^{abc}  \pi^a \pa_\mu  \pi^b  \rho_\mu^c -  g_\rho \eps^{abc}  \pa_\mu \rho_\nu^a \rho_\mu^b \rho_\nu^c 
+ {g_{\pi^4}  \over 6 v^2} \left [
\pa_\mu \pi^a \pi^a \pa_\mu \pi^b \pi^b   - \pa_\mu \pi^a  \pa_\mu \pi^a \pi^b \pi^b 
\right ]
\eeq  
where 
\beq
g_{\rho \pi \pi}  =  {\alpha \over 2}  g_\rho
\qquad  
g_{\pi^4}  = 1 - {3 \alpha  \over 4} = 1 - 3 g_{\rho\pi\pi}^2 {v^2 \over m_\rho^2}.
\eeq 
Note that the presence of the resonances automatically generates a 4-$\pi$ contact terms.
The parameter $\alpha$ sets the ratio $g_{\rho \pi \pi}/g_\rho$. 
The ``three-site model"~\cite{Chivukula:2006cg} is the special case of the above construction corresponding to $\alpha = 1$ or $g_{\rho \pi \pi}/g_\rho  = 1/2$.  
The case $\alpha = 2$ is singled out by the fact that the electroweak gauge bosons do not couple directly to $\pi\pi$~\cite{Komargodski:2010mc}.
When the same formalism is applied to describe low-energy QCD the experimentally preferred parameter range is $\alpha \approx 1.7$ and  $g_{\rho\pi\pi} \approx g_\rho \approx 6$~\cite{Barbieri:2008cc} (see Table~\ref{tab:qcd}).

Other formalisms introducing the $\rho$ meson exist, for example, $\rho$ can be defined as transforming in the adjoint of  $SU(2)_V$ and represented either by a Lorentz tensor or a vector~\cite{Ecker:1989yg, Barbieri:2008cc, Barbieri:2009tx}.
Restricting to scattering amplitude of the electroweak gauge bosons, all these formalisms are equivalent. In particular, the ``hidden gauge" formalism used in this paper can be directly translated to a formalism describing a generalized deconstruction model, as shown in Appendix~\ref{sec:deconstruction}.

We choose to describe the parameter space of our model in terms of the $g_{\rho}$ and the $g_{\rho\pi\pi}$ couplings. It might be useful however to give a dictionary connecting these parameters and other parameters used often in the literature (and defined for instance in Ref.~\cite{Ecker:1989yg})
\begin{equation}
G_{V}=m_{\rho} g_{V}=\frac{ \sqrt{\alpha} }{2} v=\frac{g_{\rho\pi\pi} v^{2}}{m_{\rho}},\ \ \ m_{\rho}^{2}=\alpha g_{\rho}^{2}v^{2}=2g_{\rho\pi\pi}g_{\rho}v^{2}.
\end{equation}

\begin{table}[t]
\begin{center}
\begin{tabular}{rl}
 $m_\rho$&$ \simeq~~ 2$ TeV\\
 $g_\rho $&$\simeq ~~6.4$\\
 $g_{\rho\pi\pi}$&$\simeq ~~5.3$\\
 $\alpha$ & $\simeq ~~1.7$
\end{tabular}
\end{center}
\caption[]{\label{tab:qcd} Comparison to QCD in the chiral limit and with $f_\pi =  v$. }
\end{table}

\section{Unitarity Constraints}\label{sec:constraints}

In this section we discuss the constraints on the maximum cutoff scale of the theory implied by perturbative unitarity of longitudinal gauge boson scattering.  
The most stringent constraints come from 2-to-2 scattering as processes with a larger number of initial or final state particles carry additional phase space suppression. 
We shall work at the leading order in the weak coupling, thus we effectively set $g,g' \to 0$ in this computation. 
Before discussing the specific amplitudes we review the general perturbative unitarity constraints. 

The unitarity constraints are customarily formulated in terms of scattering amplitudes projected into partial waves,  
\beq
\cM_{\alpha \beta}^{J}(s) = {1 \over 32 \pi} \int_{-1}^1 d(\cos \theta) \cM_{\alpha \beta} P_J(\cos \theta) 
\eeq  
where $P_J$ are the Legendre polynomials and $\alpha,\beta$ stand for 2-body initial and final states (the factor $1/\sqrt{2}$ is implicit for identical particles in the initial or final state). 
The optical theorem or the unitarity condition of the $S$ matrix relates the real and imaginary parts of the partial wave amplitude, 
\beq
\im \cM_{\alpha \beta}^J =  \sum_\gamma \cM_{\alpha \gamma}^J \sigma_\gamma  \cM_{\beta \gamma}^{J \, *}  
\eeq 
where $\sigma$ is the phase space factor: $\sigma_\alpha^2 = (1 - m_1^2/s - m_2^2/s)^2 - 4 m_1^2 m_2^2/s^2$ for $s> (m_1+m_2)^2$, and $\sigma_\alpha =0$ otherwise ($m_{1,2}$ are the masses of the two particles in the initial/final states $\alpha$ considered)~\cite{DeCurtis:2003zt}. 
If only one initial and one final state is available then one can rewrite the unitarity condition as the constraint for the amplitude to lie on the Argand circle,   
\beq
\sigma_\alpha \left ( \re  \cM_{\alpha \alpha}^J \right )^2  + \sigma_\alpha \left (\im  \cM_{\alpha \alpha} ^J -  {1 \over 2  \sigma_\alpha} \right )^2 =  {1 \over 4  \sigma_\alpha}.  
\eeq 
This leads to the usual unitarity bound  $|\re  \cM_{\alpha \alpha}^J| \leq 1/2 \sigma_\alpha$.  
For several initial and final states the condition holds for the largest eigenvalue of the matrix of amplitudes. 
At the tree level, the amplitude is real (unless the quantum width is included in the tree-level propagator), while loop corrections contribute to both the real and imaginary parts. 
Thus it is in principle possible that loop corrections bring the amplitude back inside the Argand circle.  
Defining perturbative unitarity as the condition that loop corrections to the real part of the amplitude do not exceed 50\% of the tree-level contribution leads to the unitarity condition for the tree-level part:      
\beq
\label{e.auc}
\sigma_\alpha |\cM_{\alpha \alpha}^{J, tree}| \lesssim 1  
\eeq
which we shall use in the following. With this criteria, perturbativity will be lost at $4\sqrt{\pi} v\approx 1.7$~TeV in the Standard Model without a Higgs boson. As  we could see, the above condition is arbitrary to a certain degree as it depends on assumptions about the size of the loop corrections.   However, in the cases of interest for the present paper  the tree-level amplitude will quickly grow with energy  and the scale of  unitarity violation will not depend dramatically on the numerical  coefficient  the right-hand side of \eref{auc}.

\subsection{Scattering of electroweak gauge bosons}

\begin{figure}[tb]
\vspace{1cm}
\bc
\includegraphics[width=0.9\textwidth]{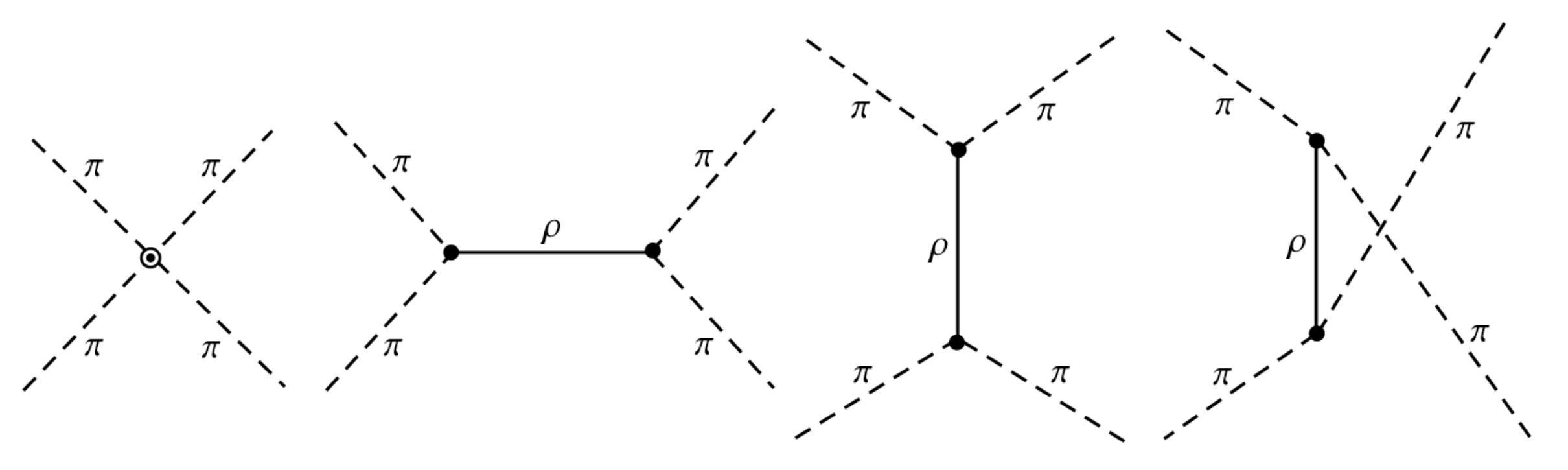}
\ec 
\caption{ \label{f.feynman_E}
\small Feynman diagrams contributing to the $\pi \pi \to \pi \pi$ scattering. 
}
\end{figure}

Although amplitudes involving $\rho$-mesons in the initial or final states provide important unitarity constraints, it is illuminating to first consider the 2-to-2  processes with only $W$ and $Z$  in the initial and final states.
We shall include the $\rho$-mesons in the next subsection.  
  
The scattering amplitudes for longitudinally polarized $W$ and $Z$ , or, equivalently, for the Goldstone bosons $\pi$ eaten by $W$ and $Z$  are given by Ref.~\cite{Bagger:1993zf}
\bea
\cM(\pi^a \pi^b \to \pi^c \pi^d) &=& \delta^{ab} \delta^{cd} M(s,t,u)  +\delta^{ac} \delta^{bd} M(t,u,s) +  \delta^{ad} \delta^{bc} M(u,s,t)
\nn 
M(s,t,u) &=& {s \over v^2} -   g_{\rho \pi \pi}^2  \left ({s - u \over t  - m_\rho^2} + {s - t \over u - m_\rho^2} + 3 {s \over m_\rho^2} \right )
\eea 
The contact terms come from the 4-pion vertex while the remaining two terms come the diagram with $\rho$ in the intermediate state. The Feynman diagrams corresponding to the $\pi \pi \to \pi \pi$ scattering are given in \fref{feynman_E}. 
Using $s+ t + u =0$ one finds that for $s \ll m_\rho^2$ this amplitude reduces, as it should, to the Higgsless SM amplitude $s/v^2$. 
This is possible only thanks to the 4-pion contact term induced by the resonances.

The most stringent unitarity constraint comes  from the  s-wave,  
\bea
\label{e.m0e}
\cM^0(\pi^a \pi^b \to \pi^c \pi^d) &= & \left [ 
\delta^{ab} \delta^{cd} - {1 \over 2}\delta^{ac} \delta^{bd} -  {1 \over 2} \delta^{ad} \delta^{bc} 
\right] \cM_{\pi\pi \to \pi\pi}^{0}(s)
\nn  
\cM_{\pi\pi \to \pi\pi}^{0}(s) &=&  {1 \over 16 \pi } \left [ { s \over v^2}   - 3 g_{\rho\pi\pi}^2 {s  \over m_\rho^2} 
-   2 g_{\rho\pi\pi}^2   + 4  g_{\rho\pi\pi}^2  \left (1+ {m_\rho^2 \over 2s} \right ) \log\left (1 + {s \over m_\rho^2} \right ) \right ] , 
\nn
\eea
that asymptotically grows as the first power of $s$ (except for $g_{\rho \pi \pi} = m_\rho/\sqrt 3 v$ corresponding to $\alpha = 4/3$, which corresponds to the $E^2$ sum rule of the 5D Higgsless models~\cite{Csaki:2003dt,SekharChivukula:2001hz} and was also observed in Ref.~\cite{Barbieri:2008cc}).  
Furthermore, the s-wave contains terms growing as $\log s$; they arise due to the poles of the $\rho$ propagator, thus their origin is IR. 
The p-wave amplitude also grows as $\co(s)$ but it always provides weaker constraints.  Higher partial waves do not grow as $\co(s)$. 
The amplitudes for scattering of the physical eigenstates are related $\cM^0$ as 
\beq
\cm^0(W_L^\pm W_L^\pm \to W_L^\pm W_L^\pm) =  -  \cM^{0}(s)  
\eeq 
\beq
 \left ( \ba{cc} 
\cm^0(W_L^+ W_L^- \to W_L^+ W_L^-  )& \cm^0(W_L^+ W_L^- \to Z_L Z_L)/\sqrt{2}
\\
\cm^0(Z_L Z_L \to W_L^+ W_L^-  )/\sqrt{2} &  \cm^0(Z_L Z_L \to Z_L Z_L )/2
\ea \right)  = 
 \cM_{\pi\pi \to \pi\pi}^{0}(s)   \left ( \ba{cc}
 {1\over 2} & {1 \over \sqrt{2}}
 \\
{1 \over  \sqrt{2}} & 0 
\ea \right)  
\eeq 
The matrix above has the eigenvalues $(1,-1/2)  \cM^{0}(s) $, thus the tree-level unitarity condition reads 
\beq
|\cM_{\pi\pi \to \pi\pi}^{0}(s) | \leq  1 
\eeq 
The maximum cut-off scale $\Lambda$ allowed by unitarity of $W$ and $Z$  scattering is determined by the lowest solution $|\cM_{\pi\pi \to \pi\pi}^{0}(\Lambda) | = 1$. 
How this maximum cut-off varies throughout the parameter space is shown in the left panel of \fref{unitarity_PU}. 
The scattering amplitudes of $W$ and $Z$  are completely  defined by 2 couplings, $g_{\rho \pi \pi}$ and $g_\rho$; the two fix $m_\rho$ via the relation 
$m_\rho^2 = 2 g_{\rho \pi \pi} g_\rho v^2$. 
We varied these two couplings in the entire perturbativity region $g_i  < 4 \pi$. 
In the white area of the plot the maximum cutoff is below $m_\rho$ which renders the set-up inconsistent.
For moderate  $g_\rho$ and $g_{\rho \pi\pi} \approx g_\rho $ the unitarity violation of $W$ and $Z$  scattering can be postponed to very large scales, up to $\sim 10$~TeV.
As pointed out in Ref.~\cite{Barbieri:2008cc}, in that region the $WW$ scattering amplitudes grow slowly because the coefficient of the $\co(s)$ term in the amplitude is slightly negative and partially cancels against the $\co(\log s)$ term.
It may be puzzling that the UV behavior of the theory relies on the $\co(\log s)$ term whose origin is IR (in particular, for the special value of $\alpha=4/3$ the $\co(s)$ term in the amplitude cancels and the UV behavior seems to depend on the $\co(\log s)$ term alone).   
However, it turns out that in that region unitarity is in fact violated at a much lower scale by the amplitudes for inelastic production of heavy resonances, as we shall see in the following.  
    
\begin{figure}[tb]
\vspace{1cm}
\centerline{
\includegraphics[width=0.47\textwidth]{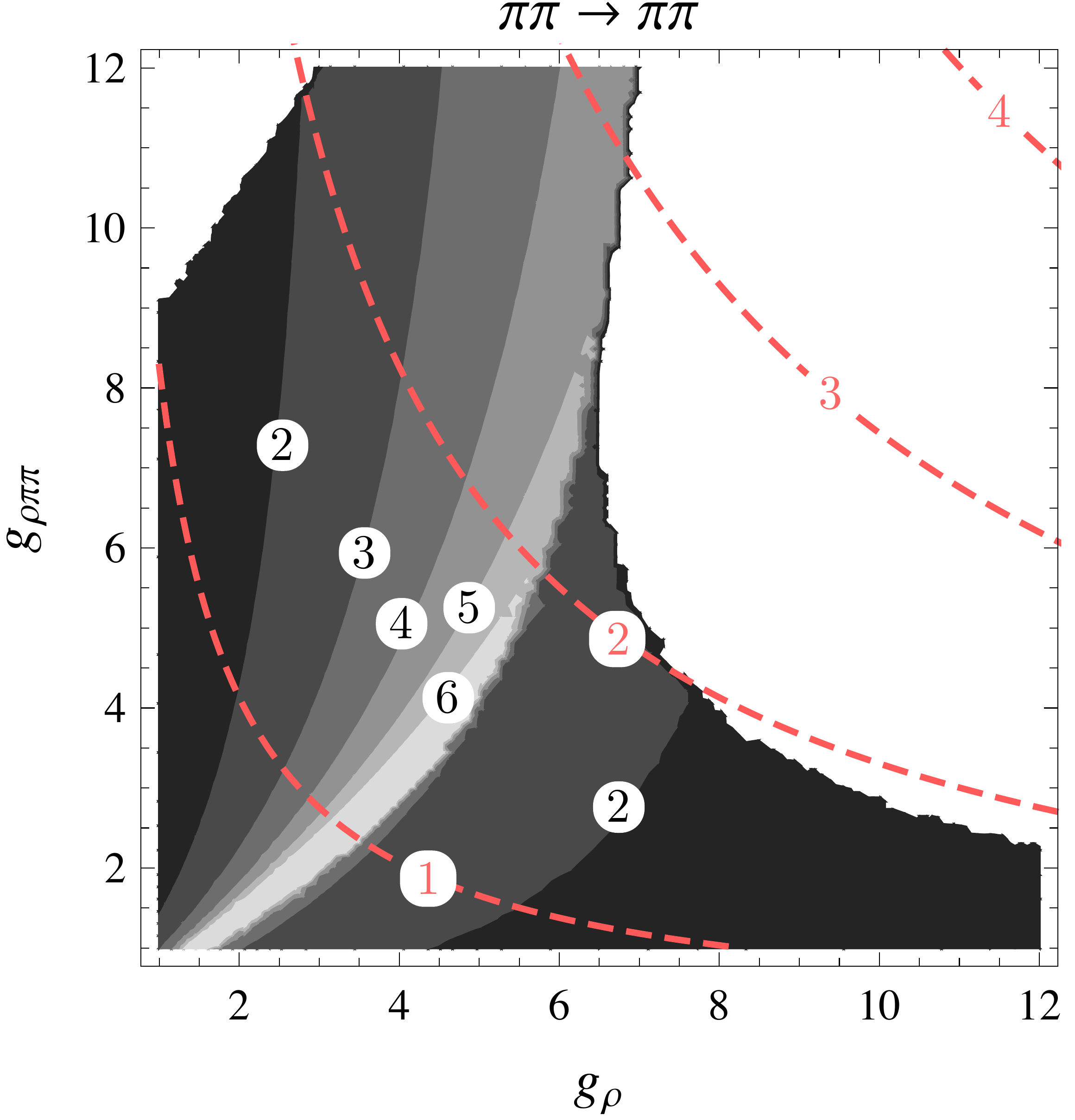}
\hspace{0.5cm} 
\includegraphics[width=0.47\textwidth]{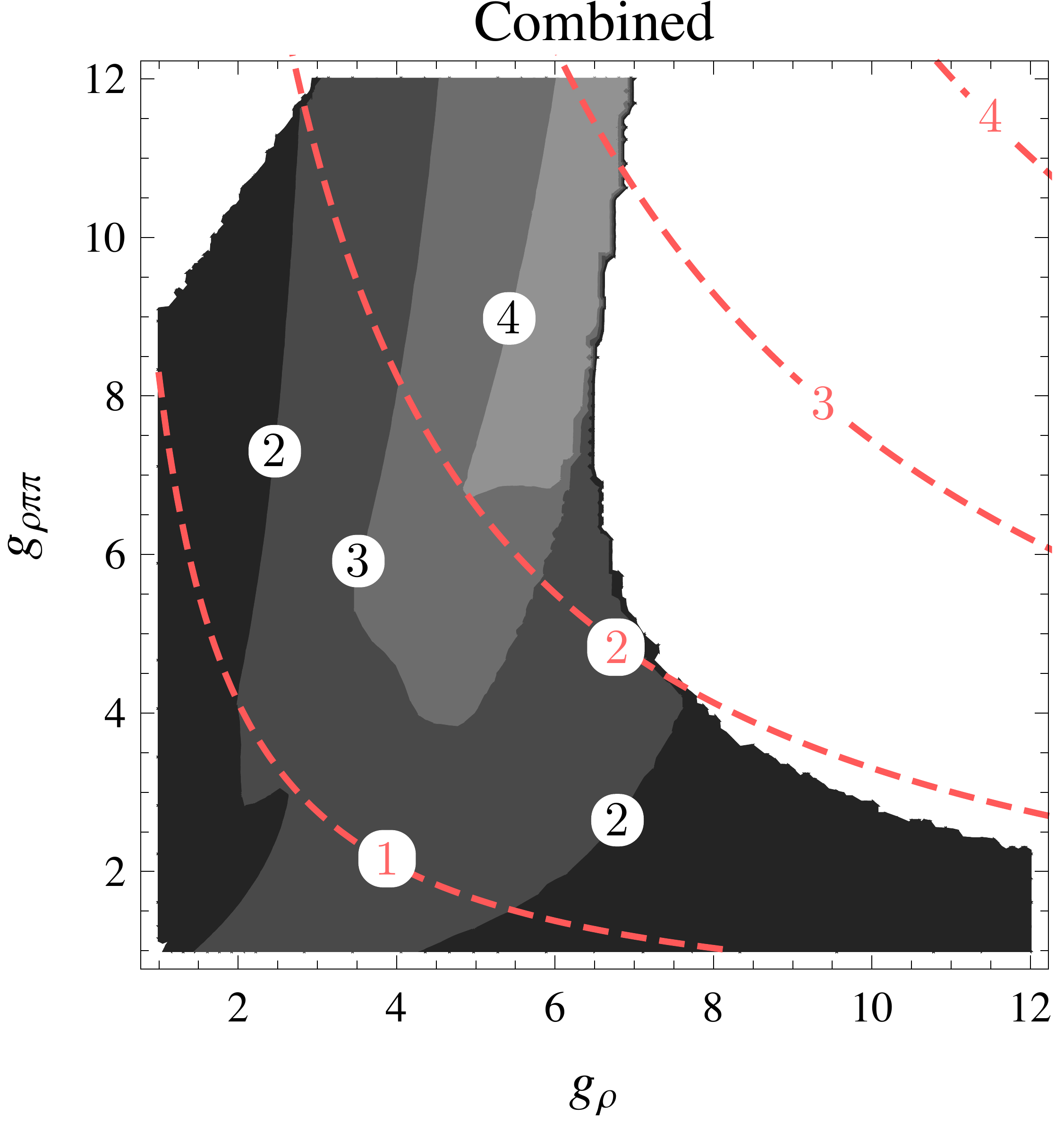}
}
\caption{ 
\label{f.unitarity_PU}
\small Contour plots for maximum cut-off scale allowed by the unitarity of the  $\pi \pi \to \pi \pi$ channel (left), 
and by all 2-to-2 channels combined (right),  overlaid with contours of constant $m_\rho/\tev$  (red dashed), 
The colored regions correspond to a cutoff scale $\Lambda$ smaller than   2, 3, 4, 5, 6~TeV (from dark to light gray).
}
\end{figure}

\subsection{Scattering into heavy resonances}

\begin{figure}[tb]
\vspace{1cm}
\bc
\includegraphics[width=0.7\textwidth]{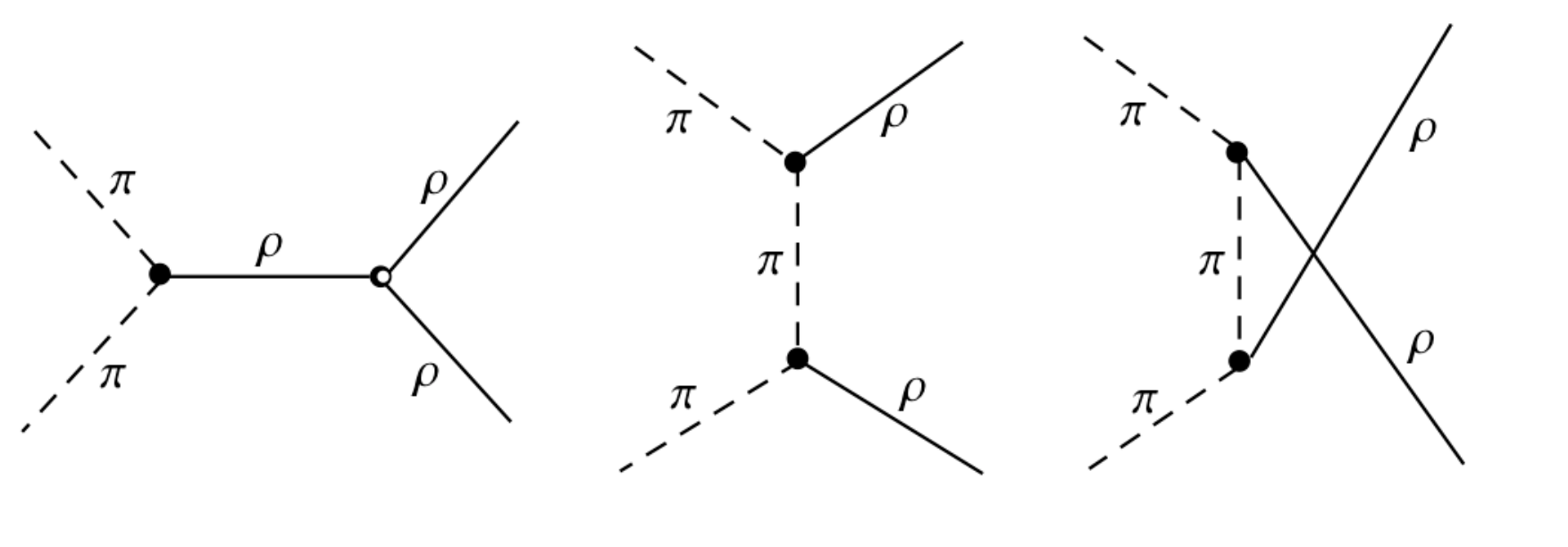}
\ec 
\bc
\includegraphics[width=0.7\textwidth]{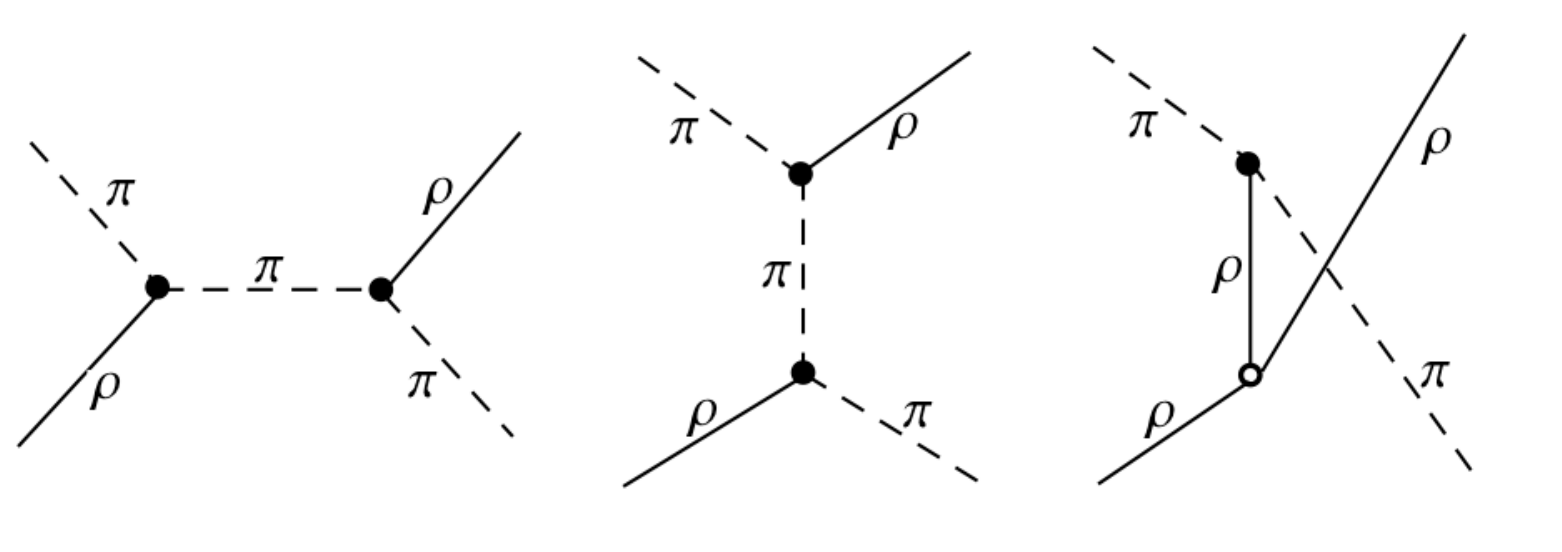}
\ec 
\caption{ 
\label{f.feynman_IE}
\small Feynman diagrams contributing to the $\pi \pi \to \rho \rho$ and $\pi \rho \to \pi \rho$  scattering. 
}
\end{figure}

Now we include the processes with $\rho$ mesons in the initial or final states and discuss their impact on unitarity violation. The Feynman digrams corresponding to these processes are given in \fref{feynman_IE}.
Consider first the inelastic production of a pair of $\rho$ mesons.  
The s-wave amplitudes for these processes are given by  
\bea 
 \label{e.m0ie}
\displaystyle
\cM^{0} (\pi^a \pi^b \to \rho_L^c \rho_L^d) &=&  
\left [ \delta^{ab} \delta^{cd} - {1\over 2}\delta^{ac} \delta^{bd} - {1\over 2}\delta^{ad} \delta^{bc} \right ] \cM^{0}_{\pi \pi \to \rho \rho}(s)
\nn  \displaystyle 
\cM^{0}_{\pi \pi \to \rho \rho}(s) &=&    {g_{\rho \pi \pi}^2 \over 16 \pi} \left ( {s \over m_\rho^2} - 2 \right ) + \co(s^{-1}) 
\eea  
Taking into account the inelastic $\rho$ production, the unitarity constraint is modified to~\cite{DeCurtis:2003zt,Papucci:2004ip}, 
\beq
|\cM_{IE}| \equiv  |\cM_{\pi\pi \to \pi\pi}^{0} | + \theta (s - 4 m_\rho^2 )\sqrt{1- 4 m_\rho^2/s} { | \cM_{\pi \pi \to \rho \rho}^{0} |^2 \over | \cM_{\pi\pi \to \pi\pi}^{0} |} \leq 1
\eeq  
where $\theta(x)$ is the Heavyside function. 
It is clear that the inelastic amplitude grows linearly with $s$, therefore it may contribute to unitarity violation at high energies.  
The coefficient of the $\co(s)$ term is always positive for arbitrary $g_{\rho \pi \pi}$. 
Therefore, the unitarity constraints from inelastic production are more stringent for large $g_{\rho \pi \pi}$, where on the other hand  the constraints from electroweak gauge boson scattering are less stringent.
Therefore there is a tension between maintaining unitarity simultaneously in all of these processes: 
the region where the electroweak scattering allows for a large cut-off will have a much lower maximum cutoff when inelastic $\rho$ production is included.  

Another constraint is provided by considering the $\rho \pi \to \rho \pi$ scattering.
The s-wave amplitudes are given by\footnote{
The full amplitude has a Coulomb singularity above $s = 2 m_\rho^2$ due to the intermediate pion going on-shell.
This singularity is an IR effect that has nothing to do with unitarity violation at high energies, and can be cured by adding an imaginary width to the initial and final $m_\rho$~\cite{Basdevant:1978tx}.    
Here we study this amplitude up to order $\co(s^0)$ where this problem does not occur. 
}  
\bea 
\label{e.m0se}
\cM^0(\pi^a \rho_L^c \to \pi^b \rho_L^d)  &=& 
  \cM_{\pi \rho \to \pi \rho} ^{0} \delta^{ab} \delta^{cd} +  \cN_{\pi \rho \to \pi \rho} ^{0} \delta^{ad} \delta^{bc}
 - (\cM_{\pi \rho \to \pi \rho} ^{0}+\cN_{\pi \rho \to \pi \rho}^{0} )\delta^{ac} \delta^{bd} 
 \nn 
\cN_{\pi \rho \to \pi \rho}^{0}    &=&  
  {g_{\rho \pi \pi}^2 \over 16 \pi} \left ( {s \over m_\rho^2} - 2 \right ) 
-   {g_\rho g_{\rho \pi \pi} \over 16 \pi}  \left ( {3 s \over 4 m_\rho^2}   -  2  +  \log(s/m_\rho^2) \right)  + \co(s^{-1}) 
 \nn 
\cM_{\pi \rho \to \pi \rho} ^{0}  &=&   
 -   {g_{\rho \pi \pi}^2 \over 32 \pi} \left ( {s \over m_\rho^2} - 2 \right ) + \co(s^{-1}) 
\eea
The s-wave amplitudes for physical process are related by 
\bea
\cm^{0}   (\pi^\pm \rho_L^0 \to \pi^\pm \rho_L^0)  = - \cm^{0}   (\pi^\pm \rho_L^\mp \to \pi^\mp \rho_L^\pm) & = &  \cM_{\pi \rho \to \pi \rho}^{0}   
\nn 
\cm^{0}   (\pi^\pm \rho_L^0 \to \pi^0 \rho_L^\pm)  = - \cm^{0}   (\pi^\pm \rho_L^\mp \to \pi^\pm \rho_L^\mp) & = &  \cN_{\pi \rho \to \pi \rho}^{0}   
\nn 
\cm^{0}   (\pi^\pm \rho_L^\pm \to \pi^\pm \rho_L^\pm) = - \cm^{0}   (\pi^\pm \rho_L^\mp \to \pi^0 \rho_L^0) & = & \cM_{\pi \rho \to \pi \rho}^{0}   + \cN_{\pi \rho \to \pi \rho}^{0}    
\eea 
Writing down the matrix in the space of these amplitudes one finds that the unitarity constraints read 
\beq  
2 (1- m_\rho^2/s)\left |\cM_{\pi \rho \to \pi \rho}^{0}   + \cN_{\pi \rho \to \pi \rho}^{0} \right | \leq 1 \qquad (1- m_\rho^2/s) \left |\cM_{\pi \rho \to \pi \rho}^{0}   - \cN_{\pi \rho \to \pi \rho}^{0} \right |\leq 1  
\eeq 
The amplitudes again grow linearly with $s$ and may lead to unitarity violation. 
Furthermore, the energy threshold for $\pi \rho$ scattering is at $s = m_\rho^2$, compared to $4 m_\rho^2$ for the inelastic $\rho$ production. 
We find that in certain regions of the parameter space the unitarity is first lost in the  $\pi \rho$ scattering amplitude, before it is lost in the electroweak and inelastic $\rho$ amplitudes.

The maximal allowed cut-off in the entire parameter space is displayed in the right panel of \fref{unitarity_PU}.  
The region of very high cut-off shown in the left panel disappears once the channels with the $\rho$ mesons are taken into account. 
Nevertheless, viable regions of the parameter space exist with $\Lambda$ as high as  $\sim 4$~TeV and  $m_\rho \sim 2$~TeV. 

In order to illustrate the behavior of all the considered scattering amplitudes more clearly we have plotted the energy of perturbative unitarity violation as a function of $g_{\rho\pi\pi}$ for two specific values of $m_{\rho}$ (\fref{unitarity_PB_s}). 
For small values of $g_{\rho \pi\pi}$ the longitudinal $W$ and $Z$  scattering amplitudes grow monotonically with $s$. 
The maximal possible value of the cutoff in the elastic $\pi\pi\rightarrow\pi\pi$ channel is obtained for a specific ``critical" value of $g_{\rho\pi\pi}$ above which the amplitude does not behave monotonically in $s$. Starting from that value of $g_{\rho\pi\pi}$ the $\pi\pi\rightarrow\pi\pi$ amplitude as a function of $s$ starts diminishing before it reaches $1$ and violates the unitarity constraint after it becomes negative (a ``turnaround" is possible). The most optimal region for prolonging unitarity corresponds to the values of $g_{\rho\pi\pi}$ close to this ``critical" value. One can observe that taking into account the amplitudes involving the $\rho$ mesons drastically lowers the cutoff scale in this optimal region of $g_{\rho\pi\pi}$. Plotting the maximal cutoff as a function of $m_{\rho}$ (\fref{unitarity_max}) one can see that the unitarity constraint from the $\pi\pi\rightarrow\rho\rho$ channel dominates for low values of $m_{\rho}\sim 1-2$~TeV, placing the cutoff almost immediately above $2m_{\rho}$. For intermediate values of $m_{\rho}\sim 2.5$~TeV the most stringent constraint comes from the $\pi\rho\rightarrow\pi\rho$ channel, bringing the cutoff below $2m_{\rho}$. For large values of $m_{\rho}\sim 3$~TeV the $\pi\rho\rightarrow\pi\rho$ channel constraint becomes weaker while the elastic $\pi\pi\rightarrow\pi\pi$ channel determines the cutoff (which is below $2m_{\rho}$, so the $\pi\pi\rightarrow\rho\rho$ channel does not constrain it due to kinematical reasons). As $m_{\rho}$ increases the optimal ``turnaround" region shifts to larger and larger values of $g_{\rho\pi\pi}$ and at some point it becomes unreachable because of the perturbativity constraint. This results in a drastic decrease of the maximal cutoff as a function of $m_{\rho}$ for $m_{\rho}\sim 3.2$~TeV.

\begin{figure}[h!]
\vspace{.5cm}
\centerline{
\includegraphics[width=0.47\textwidth]{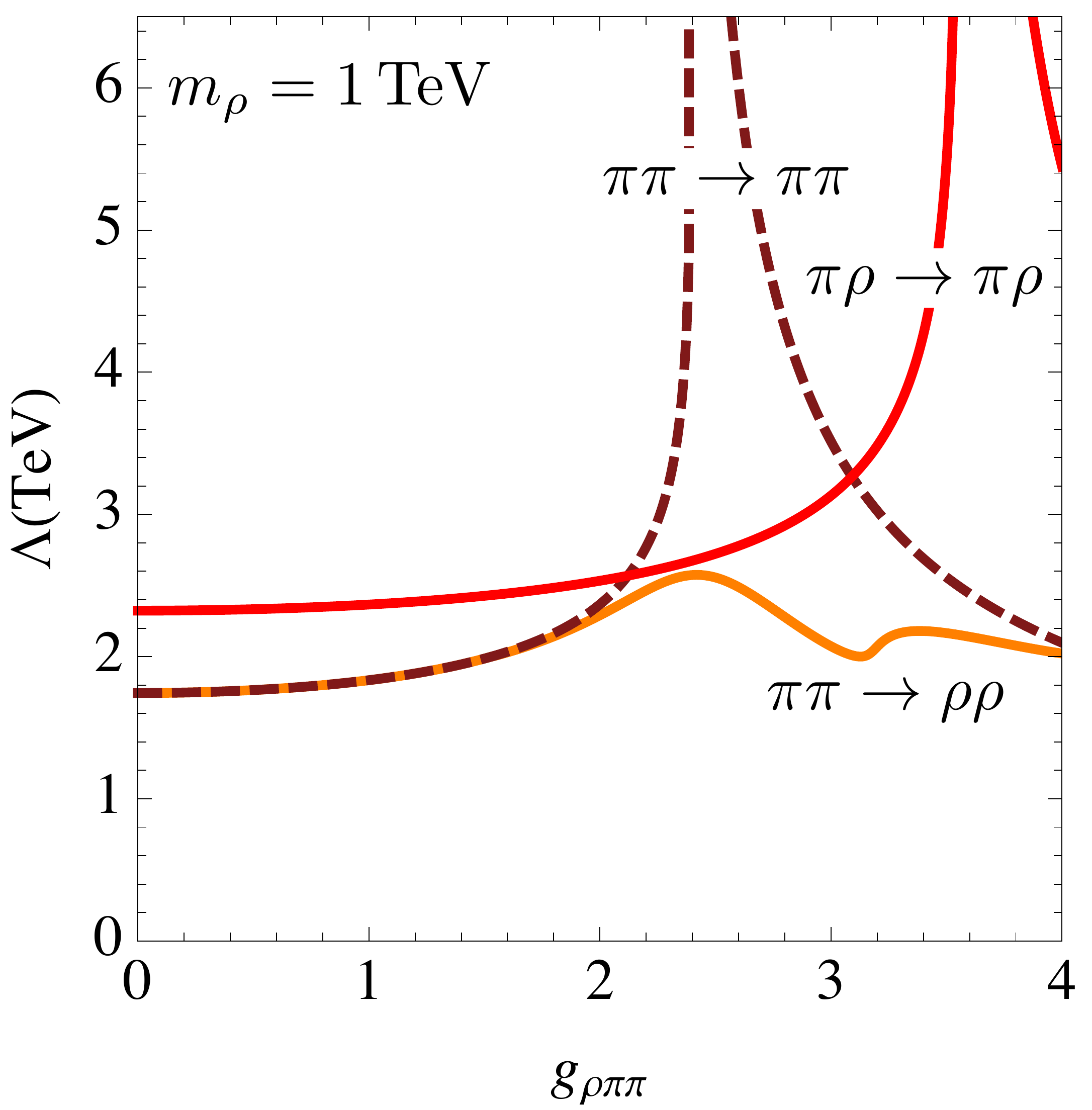}
\hspace{.5cm} 
\includegraphics[width=0.47\textwidth]{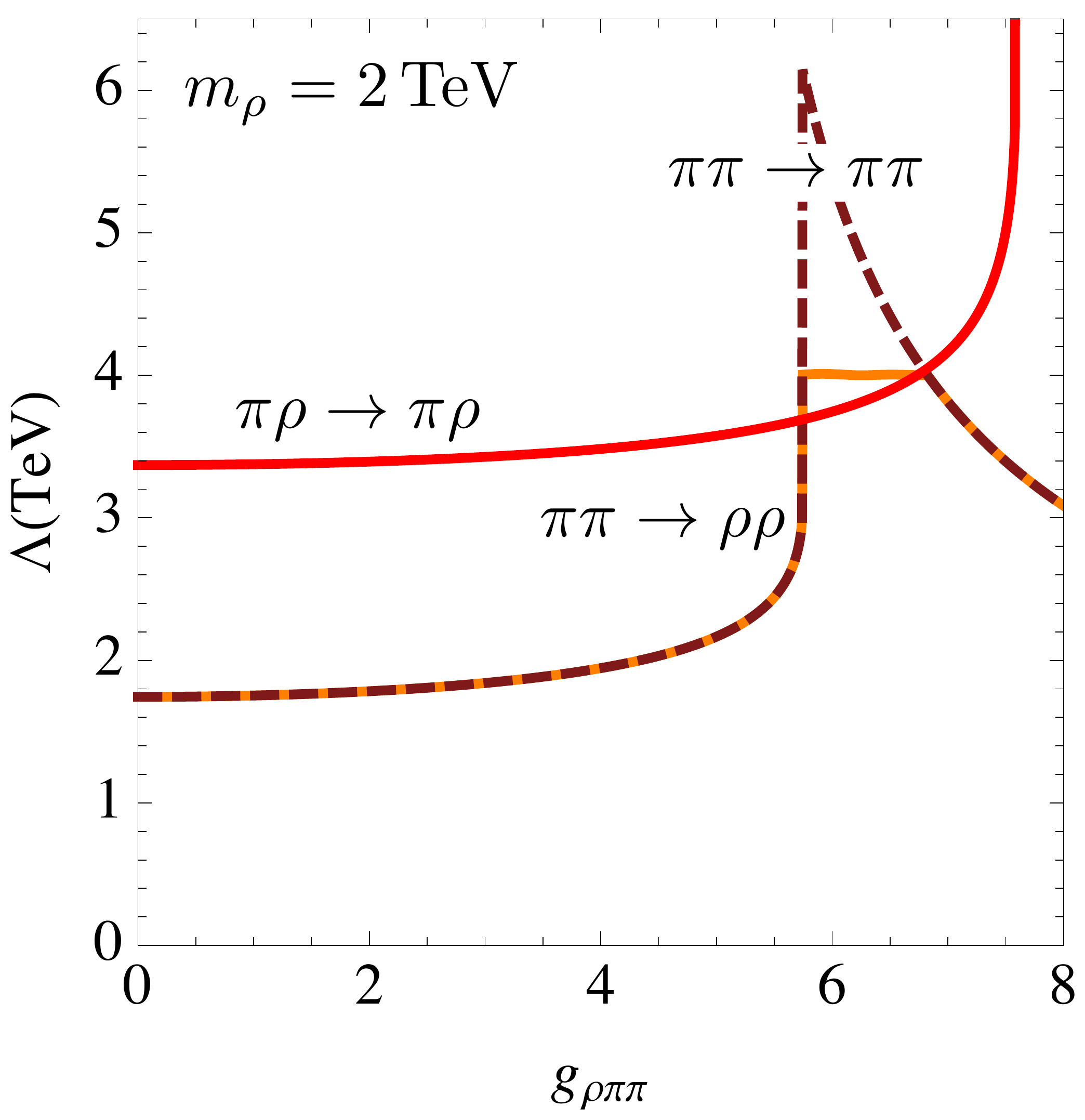}
}
\caption{
\label{f.unitarity_PB_s}
\small The maximal cutoff scale allowed by unitarity in the $\pi\pi \to \pi \pi$ (dashed brown),  $\pi\pi \to \rho \rho$ (orange), and $\pi \rho \to \pi \rho$ (red) channels 
for  $m_\rho  = 1$~TeV (left) and $m_\rho  = 2$~TeV (right) as a function of the $\rho\pi\pi$ coupling. The optimal for unitarity ``turnaround" region in the $\pi\pi \to \pi \pi$ amplitude (explained in the text) is clearly distinguishable. Note that in the limit $g_{\rho \pi\pi}\to0$, we recover the cutoff of the SM without the Higgs boson, 1.7~TeV.
}
\end{figure}

\begin{figure}[h!]
\vspace{.5cm}
\centerline{
\includegraphics[width=0.47\textwidth]{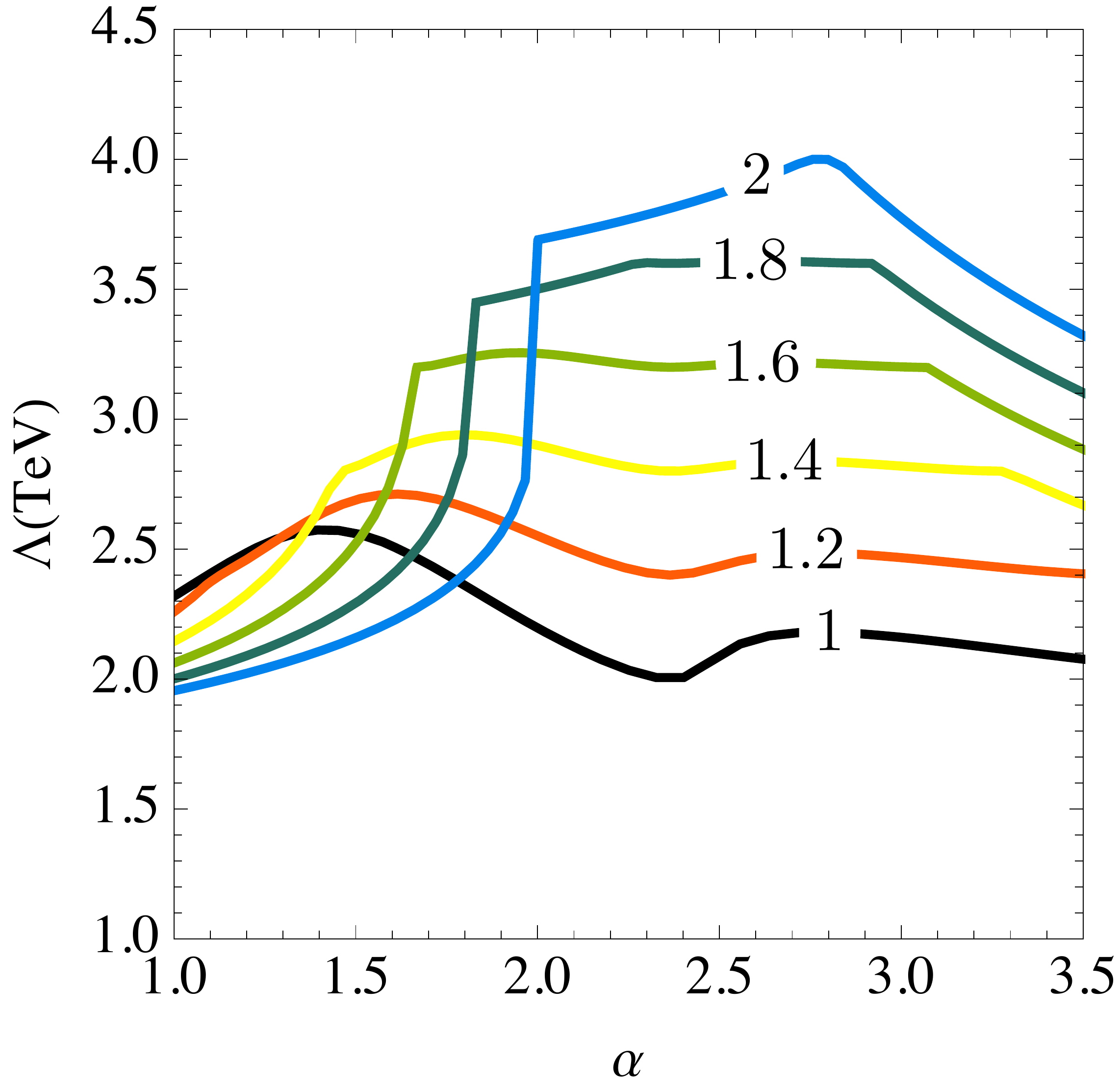}
\hspace{0.5cm} 
\includegraphics[width=0.44\textwidth]{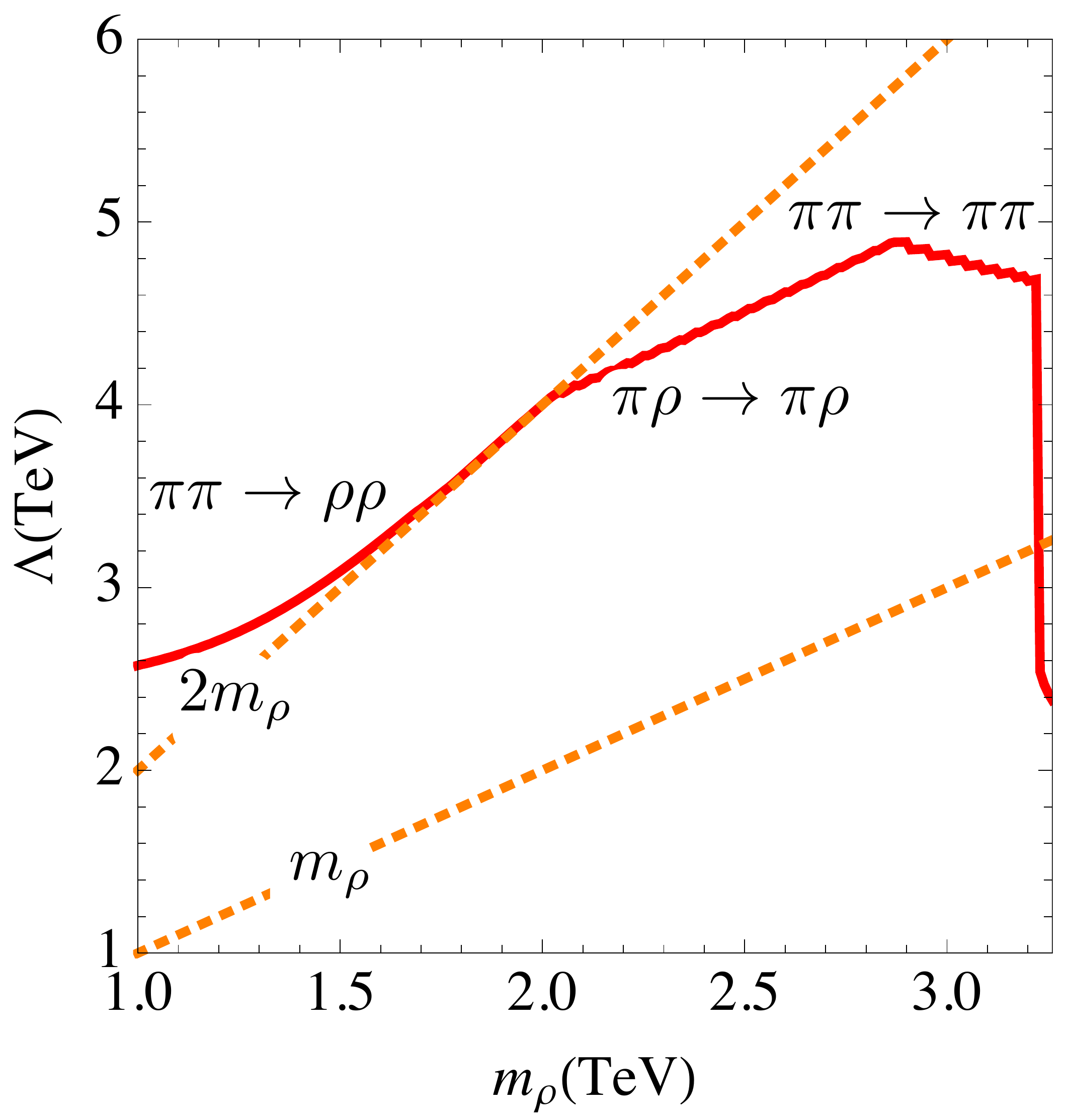}
}
\caption{
\label{f.unitarity_max}
\small The maximal cutoff scale allowed by unitarity (summing all considered channels) for several values of $m_\rho$ [TeV] as a function of $\alpha$ (left)  and the maximal possible value of the cutoff as a function of $m_{\rho}$ (right). Again, at fixed $m_\rho$, in the limit $\alpha\to 0$ we recover the cutoff of the SM without the Higgs boson. 
}
\end{figure}

\section{Phenomenology of resonances }

The resonances emerging from a strong sector which could be responsible for the breaking of the electroweak symmetry have been under scrutiny since the pioneering study of Ref.~\cite{Bagger:1993zf}. After the disgrace of technicolor models as a result of the LEP precision measurements, there has been revival of interest in strong electroweak symmetry breaking models thanks to their duality with perturbative models built with compactified or deconstructed extra dimensions. The LHC (and ILC) phenomenology of these Higgsless models gave rise to many studies~\cite{higgslesspheno}. In the past few years, there has been some interest  on more minimal models~\cite{Barbieri:2008cc, Chivukula:2006cg, Accomando:2008jh} inspired by the QCD chiral Lagrangian which are also reminiscent of the original BESS models~\cite{Casalbuoni:1985kq}. There is an abundant literature on the phenomenology of these strong EW resonances (see e.g. Refs.~\cite{heavyvectorsLHC}). We review the main results of these analyses and we extend them by a study of the LHC discovery potential.

\begin{figure}[h]
\vspace{0.5cm}
\centerline{
\includegraphics[width=0.47 \textwidth]{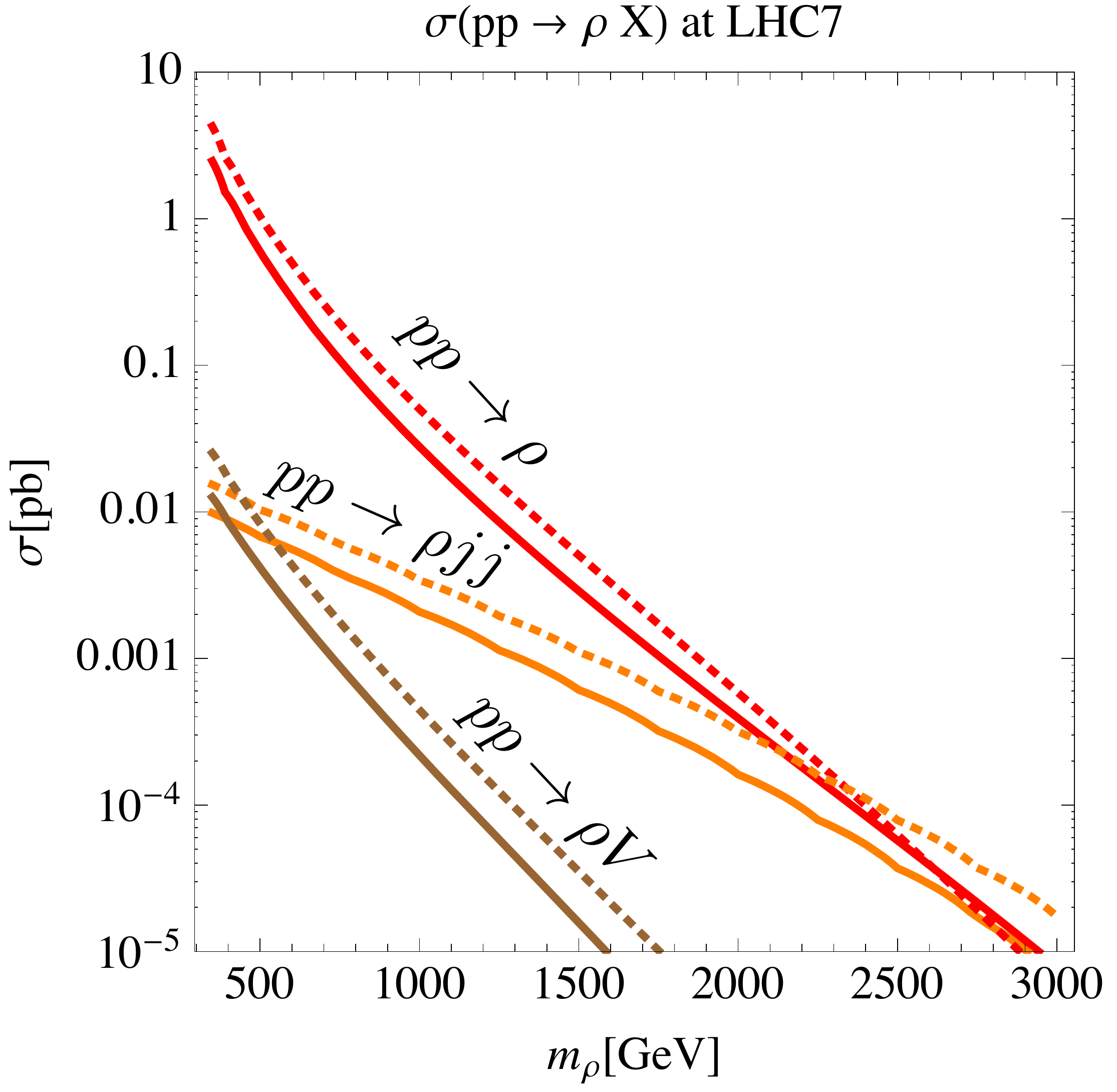}
\hspace{0.5cm}
\includegraphics[width=0.47 \textwidth]{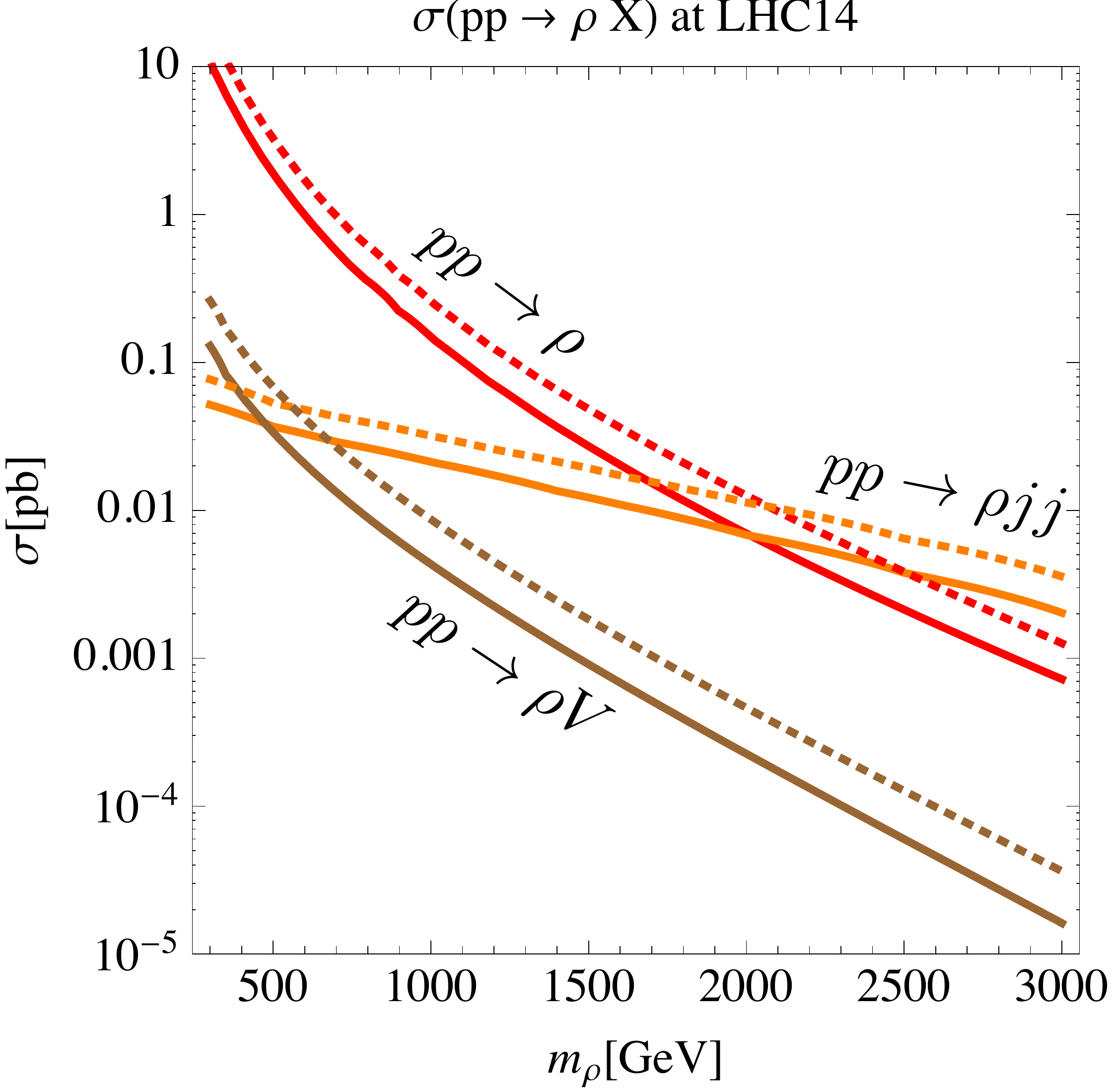}
}
\caption{
\label{f.sigmaLHC7}
\small Cross section for the production of a single neutral (solid) and charged (dashed) resonance at the LHC  with $\sqrt{s} = 7$~TeV (on the left) and $\sqrt{s} = 14$~TeV (on the right) in the  Drell-Yan (red), VBF (orange) and $\rho$-strahlung (brown) channels. 
We set $g_\rho = 4$;
for different coupling these cross section scale as $1/g_{\rho}^2$. 
}
\end{figure}

\subsection{Leading interactions}

In the following we assume that the SM quarks and leptons are fundamental, that is to say, they couple to the heavy resonances only via mixing of the latter with  the SM gauge bosons.\footnote{In specific models some fermions, especially the 3rd generation quarks, may have a large composite component and therefore a larger coupling to the heavy resonances. 
This would lead to a sizable branching fraction for the decay of the resonances into these fermions, see e.g. Ref.~\cite{agasheEW}. Alternatively, a suppressed coupling can also be achieved and which can improve electroweak precision fits~\cite{Cacciapaglia:2004rb}.
 }
This mixing arises due to non-diagonal entries in the gauge boson mass matrix implied by the lagrangian \eref{LeadingLagrangianPC}. 
At the leading order in $1/g_\rho$ the mass eigenstates are reached by the rotation of the SM gauge bosons (see Appendix~\ref{sec:eigenstates})
\bea
W_\mu^\pm &\to&  W_\mu^\pm   - {g \over 2 g_\rho}   \rho_\mu^\pm , \nn  
Z_\mu &\to & Z_\mu - { g^2 - g'{}^2 \over 2 g_\rho \sqrt{g^2+  g'{}^2} }\rho_\mu^0 ,  \nn 
A_\mu &\to&  A_\mu - {e \over 2 g_\rho}  \rho_\mu^0 , 
\eea 
and the corresponding rotation  of $\rho$. 
As a result, the heavy mass eigenstates $\rho^0,\rho^\pm$ couple to the SM fermions,  
\beq
\label{e.rhoff}
-   {g^2 \over 2 \sqrt 2 g_\rho}  \rho_\mu^\pm  \ov f_L \gamma_\mu T^\pm f_L    -   {1 \over 2 g_\rho}  \rho_\mu^0   \ov f \gamma_\mu \left( (g^2 - g'{}^2) T^3  + g'^2 Q  \right)   f  . 
\eeq 
where $T^\pm = (\sigma^1 \pm i \sigma^2)/2$. 

Furthermore, the SM gauge boson self interactions  after the rotation produce the couplings of $\rho$ to the electroweak gauge bosons. 
 In particular, the cubic gauge boson vertices with one $\rho$ are given by 
\beq 
\label{e.rhovv}
\! \! \! \!
 - {g^2 \over 4 g_\rho} \left (\pa_\mu W_\nu^+ W_\mu^- - \pa_\mu W_\nu^- W_\mu^+  \right )\rho_\nu^0   
-   {g \sqrt{g^2 + g'{}^2} \over 4 g_\rho} \left \{ (\pa_\mu W_\nu^- Z_\mu - \pa_\mu Z_\nu W_\mu^-  )\rho_\nu^+ 
+  \hc  \right \} 
+  \dots  
\eeq  
where the dots stand for cyclic permutations of the fields in each vertex. 

\subsection{Decays} 

The cubic gauge vertices in \eref{rhovv} induce the dominant decay of $\rho$ is into a pair of longitudinally polarized electroweak gauge bosons. 
The leading order decay widths can be computed using the Goldstone boson equivalence theorem,  
\beq
\Gamma(\rho^0 \to  W^+ W^-)  \approx   \Gamma(\rho^\pm \to  Z W^\pm) \approx 
 {m_\rho g_{\rho \pi \pi}^2 \over 48 \pi}  =  {m_\rho^5 \over 192 \pi g_\rho^2 v^4}  \, . 
\eeq
In our numerical analysis below we use the full $\rho \to VV$ matrix element that also takes into account decays into transversely polarized gauge bosons. 
These correct the leading order widths by $\sim 50$\% for $m_\rho \sim 350$~GeV, and by $\sim 10$\% for $m_\rho \sim 1$~TeV. 
In \eref{rhovv} the charged resonances  couple to $W Z$ and not   to $W \gamma$. 
This is a consequence of our assumption that the strength of the $\rho^3$ vertex in the original lagrangian  is set by the hidden $SU(2)$ gauge coupling $g_\rho$. 
Departure from the gauge coupling, $g_{3\rho} = g_\rho + \delta$, would result in the $\rho W \gamma$ vertex suppressed by $\delta g^2/g_\rho^2$ which would allow for subleading decays $\rho^\pm \to W^\pm \gamma$, as studied in Ref.~\cite{Grojean:2011vu}.    

The heavy resonances also decay to the SM fermions via the couplings in \eref{rhoff},   
however,  these decays  are strongly suppressed in the interesting parameter space (for $m_\rho \gg 2 m_W$). 
For example, the leptonic branching fractions are given by 
\beq 
{\rm Br}(\rho^\pm \to e^\pm \nu) \approx   2 {\rm Br} (\rho^0 \to e ^+ e^-) \approx  {16 m_W^4 \over  m_\rho^4 }  
\eeq
For $m_\rho \sim$~TeV  this is already less than $10^{-3}$. 
Conversely, the branching fraction into the electroweak gauge bosons is practically equal to 1 throughout the interesting parameter space.  
Thus, the main discovery channel at the Tevatron and LHC is the search for resonant production of $W^+ W^-$ and $W^\pm Z$ pairs.  

\subsection{Production and direct searches}

In hadron colliders, the resonances are produced mainly via the following processes:
\ben 
\item Drell-Yan (DY),  $q \bar q \to \rho$: a quark-antiquark collision produces a single $\rho$ thanks to the vertices contained in \eref{rhoff}  
\item Vector boson fusion (VBF), $VV \to \rho$:  both incoming quarks or antiquarks emit $W$ or $Z$  boson who collide and produce $\rho$ via the vertices in \eref{rhovv}.
This leads to the production of a single resonance in association with 2 light spectator jets in the forward direction. 
\item  $\rho-$strahlung, $V \to \rho V$:  a quark-antiquark collision produces an off-shell $W$ or $Z$  who emits  $\rho$ via the vertices in \eref{rhovv}.   
This leads to the production of a single resonance in association with an electroweak gauge boson.  
\een  
The cross section depends on  $m_\rho$ via the parton distribution functions.  
Furthermore, since the coupling of the resonances to the SM is suppressed by $1/g_\rho$ for a fixed $m_\rho$ the cross section for all the above processes scale as $1/g_{\rho}^2$. 
In \fref{sigmaLHC7} we plotted the cross sections for the three channels above at the LHC with $\sqrt{s} = 7$~TeV and $\sqrt{s} = 14$~TeV. 
The Drell-Yan process dominates in most of the parameter space.  
The VBF is suppressed by the 3-body final state phase space.  
However it becomes important for very heavy resonances, $m_\rho \simgt 2$~TeV  because this process, unlike the two others, can be initiated by a quark-quark collision, and the quark-quark luminosity at the LHC decreases less rapidly than the quark-antiquark one. 
The $\rho$-strahlung cross section is always down by approximately 2 orders of magnitude compared to Drell-Yan.  
 
In \fref{paramspace} we plot the contours of the inclusive $\rho$-production cross section on top of the parameter space allowed by perturbative unitarity. 
We also estimate the impact of the existing collider searches on the allowed parameter space. 
Currently, the best limits come from the CMS search for $WZ$ resonant production \cite{cmsWZ} which supersede the earlier Tevatron constraints ~\cite{Abazov:2010dj}. 
Taken at face value, CMS  excludes only the corner of the parameter space corresponding  $m_\rho < 900$~GeV because the  limits presented in Ref.  \cite{cmsWZ} do not extend above $900$~GeV. However, since no $WZ$ events with invariant masses larger than $900$~GeV are observed in  Ref. \cite{cmsWZ}, it should be possible to extend the limits to higher $m_\rho$  as long as the efficiency for detecting $WZ$  pairs does not drop abruptly above $m_{WZ} = 900$~GeV. 
Assuming this efficiency remains roughly constant would imply  the limit  $\sigma(pp\to \rho^\pm) {\rm Br}(\rho^\pm \to W^\pm Z) \lesssim 0.01$ pb in which case CMS excludes resonance masses up to 1-1.5 TeV, depending on the magnitude of $g_{\rho}$.  
LHC searches for $Z'$ and $W'$ in the dilepton channel (see for example Ref.~\cite{atlaszprime}) are far less sensitive due to a small leptonic branching fraction of the $\rho$. 
  
Let us now sketch the discovery potential for VBF and DY production of the resonances at the LHC. We list some benchmark values for the cross sections in Table~\ref{tab:xsec14}.
For VBF we use a recent ATLAS study~\cite{ATLASwwres}. The analysis uses updated techniques to deal with boosted W's~\cite{butterworth} and includes a complete
modeling of detector effects. Even though the definition of the signal is slightly different - they use a 
unitarization scheme whose physical meaning is obscure - we can draw some conclusions about the reach at high integrated luminosity.

The efficiency$\times$acceptance ($\epsilon\times A$) for the most promising semi-leptonic $qqWW$ channel 
is  quite low and the backgrounds after cuts remain sizable ($\sim 0.5$ fb). Assuming the quoted ($\epsilon\times A$) applies to our case, we find that a discovery of a 1 TeV (2 TeV) resonance with $g_\rho=4$ ($g_\rho=6$)
requires about 75~fb$^{-1}$ (2.5~ab$^{-1}$). Note, that for the higher mass reach we have  assumed
that the backgrounds after cuts are of similar size than for low mass, which is likely too pessimistic.
We find, that especially in the case of strong coupling or high mass resonances, VBF is clearly a challenging channel and improvements of the analysis would be very welcome. The CMS collaboration is currently studying similar channels and should present some expected reaches soon. 

The case of DY is more promising but also more model-dependent since the coupling to fermions could in principle 
be very different from the minimal coupling through mixing. In the case of partial compositeness for example, the strength of the coupling to fermions is linked to the mass of the fermion and can be $\sim g_\rho$ for the 3rd generation (compared to $g/g_\rho$ from mixing), see e.g.~Ref.~\cite{agasheEW}. Alternatively, for a flavor invariant strong sector the coupling to the light generations can be very large~\cite{MFVCPH}. Both of these possibilities would lead to different cross-sections and  imply  very different dominant final states. 
Keeping this in mind, let us from now on focus on the minimal case. 

DY production has been studied in Ref.~\cite{Cata:2009iy} and we agree with their results for the cross-section (see also Refs.~\cite{heavyvectorsLHC}). A recent analysis Ref.~\cite{katzWW}, explores the potential of 
jet substructure methods for discovering a $Z'$ decaying to $WW$. The authors find the semi-leptonic channel to be the most promising and show that a good ($\epsilon\times A$) can be achieved. Using the quoted discovery reaches for the signal cross-sections, we estimate that a neutral $\rho$ with a mass of 1~TeV (2~TeV)  and a coupling $g_\rho =4$ ($g_\rho=6$), can be discovered at LHC14 after accumulating about 
5~fb$^{-1}$ (85~fb$^{-1}$).

\begin{table}[t]
\begin{center}
\begin{tabular}{ccc|cccccc}
 $g_\rho$ & $m_\rho$ [TeV]  &  $\Gamma/m_\rho$ & \text{DY [fb]} & \text{VBF [fb]} & $\rho V$ [fb] & \text{DY$^\pm$ [fb]} & \text{VBF$^\pm$ [fb]} & $\rho^\pm V$ [fb]  \\\hline
  4 & 1 & 0.031 & 146 & 21 & 4.3 & 255 & 32 & 8.7 \\
 4 & 1.5 & 0.15 & 27 & 12 & 0.91 & 48 & 19 & 1.8 \\
 4 & 2 & 0.46 & 7.0 & 6.8 & 0.23 & 12 & 11 & 0.46 \\
 6 & 1 & 0.014 & 65 & 9.4 & 1.92 & 114 & 14 & 3.8 \\
 6 & 1.5 & 0.066 & 12 & 5.4 & 0.40 & 21 & 8.6 & 0.81 \\
 6 & 2 & 0.21 & 3.1 & 3.0 & 0.10 & 5.6 & 5.0 & 0.21\\
 6 & 2.5 & 0.50 & 0.95 & 1.7 & 0.027 & 1.7 & 2.9 & 0.056\end{tabular}
\end{center}\caption {Benchmark values of the production cross-sections for neutral and 
charged resonances at the LHC with $\sqrt{s} = 14$~TeV.  }
\label{tab:xsec14}
\end{table}
  
In conclusion, we find that in the interesting mass range and for $g_\rho=4$, DY produced resonances should be discoverable if they are not too broad, whereas VBF requires very large integrated luminosities (or an improved analysis). For larger coupling $g_\rho=6$, the cross-sections are smaller due to the reduced mixing ($\sigma \sim 1/g_\rho^2$) and the required integrated luminosities increase by roughly a factor of four. Further, the heavier the resonances the broader they are, complicating the searches even more.
One easily enters the asymptotic regime 
of LHC and a discovery of the degree of freedom unitarizing $W_L W_L$ scattering can not be guaranteed. This would truly constitute a nightmare scenario.

\begin{figure}[h]
\vspace{0.5cm}
\centerline{
\includegraphics[width=0.47 \textwidth]{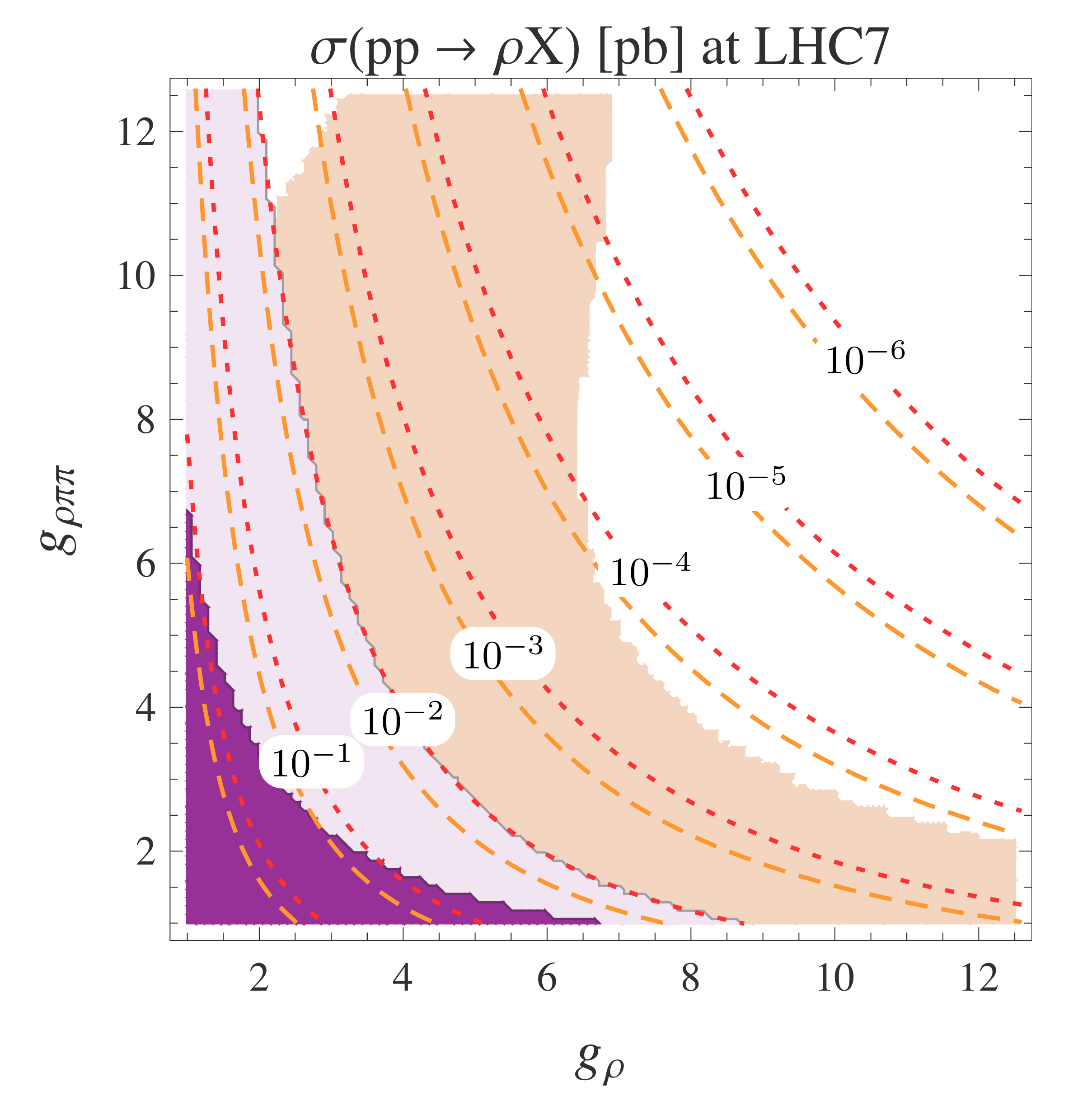}
\hspace{0.5cm} 
\includegraphics[width=0.46 \textwidth]{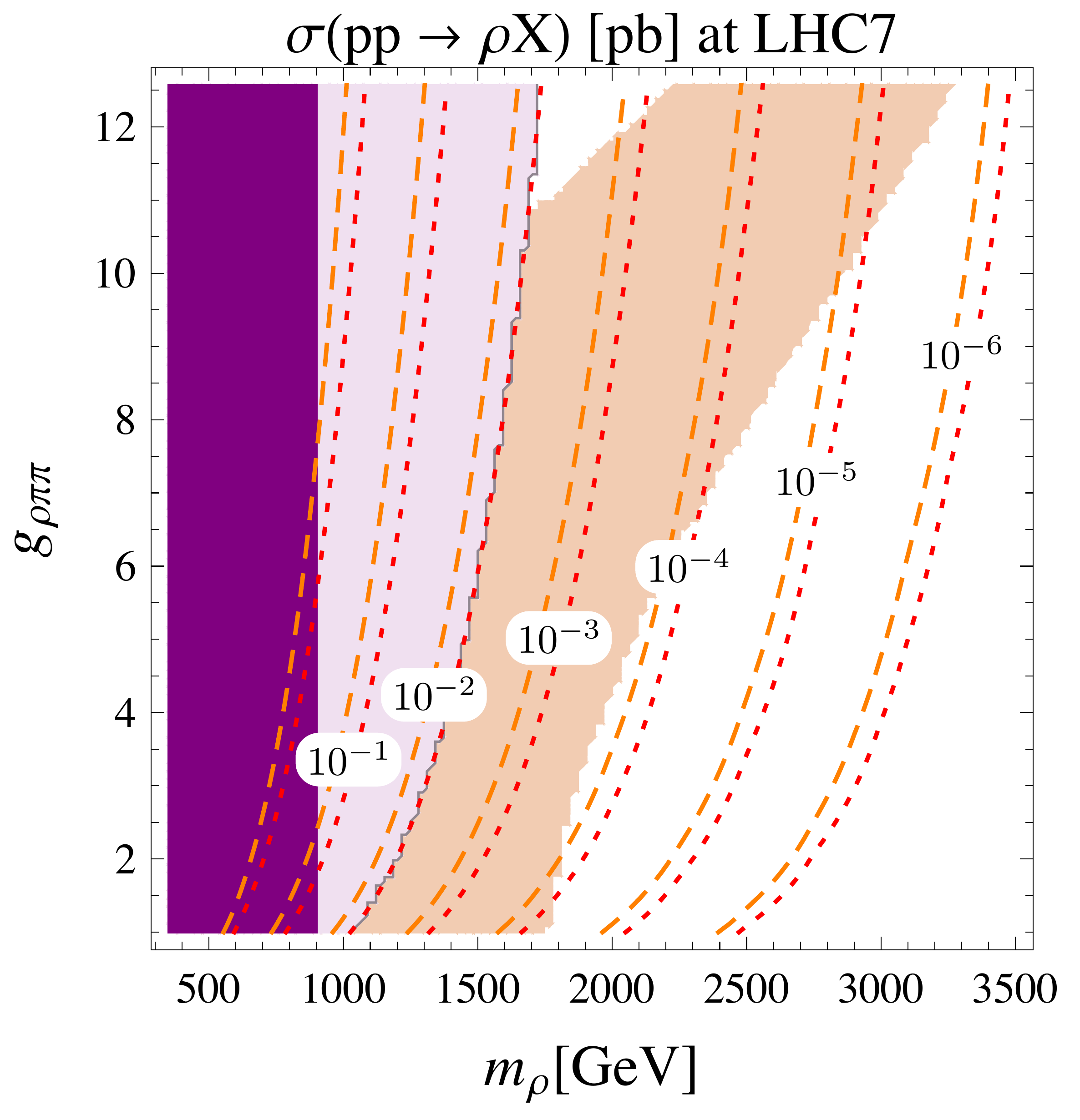}
}
\caption{
\label{f.paramspace}
\small The viable parameter space of our model in the $g_\rho - g_{\rho \pi\pi}$ plane (left) and $m_\rho - g_{\rho \pi\pi}$ plane  (right). 
We give the contours of the total cross section for the inclusive production of  $\rho^0$,  $\rho^\pm$ (dashed, dotted) at the LHC with $\sqrt{s}  = 7$~TeV which we computed at tree level using the MSTW 2008 PDFs~\cite{Martin:2009iq}. 
The Drell-Yan cross section is computed in the narrow width approximation which becomes less reliable for  $g_{\rho \pi \pi} \simgt 6$. 
The light orange area is allowed by the unitarity constraints on longitudinal gauge boson scattering in elastic and inelastic channels.
The CMS search for $WZ$ resonant production \cite{cmsWZ}  excludes the region with $m_\rho \leq 900$~GeV (deep purple). 
We  also show the approximate exclusion range of the CMS search if their limits are extrapolated to $m_\rho > 900$~GeV (light purple).  
}
\end{figure}

\subsection{Indirect constraints}

Below the scale $m_\rho$ one can integrate out the heavy resonance so as to obtain the effective theory describing the SM gauge and fermion degrees of freedom. 
At the tree level the procedure amounts to solving the equations of motion for $\rho$ and plugging the solution back to the lagrangian.  
This effective theory includes the SM lagrangian (without the Higgs) and  oblique corrections~\cite{Peskin:1991sw, Barbieri:2004qk}
to the SM gauge boson propagators.    
The $T$ parameter is zero at the tree level thanks to the custodial symmetry imposed on the strong sector.  
The $W$ and $Y$ parameters of Ref.~\cite{Barbieri:2004qk} are suppressed by $g^4/g_{\rho}^4$ and are not important. 
For the $S$ parameter one finds 
\beq
\label{e.sparam}
\Delta S  = {4 \pi \over g_\rho^2} 
\eeq 
This contribution is much larger than the LEP limit of $S \simlt 0.2$ unless  $g_\rho$ is near the perturbativity limit.  
However one can envisage the strong sector producing additional contributions to $S$ that cancel against \eref{sparam}~\cite{Barbieri:2008cc}.     
One possibility is adding an axial resonance with appropriately tuned mass and couplings.  
Furthermore,  in a setup with only a vector resonance, the symmetries of the strong sector admit the following $\co(p^4)$ operator:  
\beq
\label{e.opeps}
 - {\eps \over 16 g_\rho} \tr \left \{  [g  \xi_{L}^\dagger  L_{\mu\nu}^a \sigma^a  \xi_{L}  + g'   \xi_{R}^\dagger  B_{\mu\nu} \sigma^3  \xi_{R}  ] \rho_{\mu \nu} 
\right \} 
\eeq 
Upon integrating out $\rho$, this contributes  $\Delta S =   4 \pi \eps/g_\rho^2$ and choosing $\eps < 0$ one can tune away the $S$ parameter.\footnote{Adding the operator \eref{opeps} is equivalent to choosing  $F_V \neq 2 G_V$ in Ref.~\cite{Barbieri:2008cc}.} .  

Integrating $\rho$ at the one-loop level one obtains contributions to the $T$ parameter. 
For $\eps = 0$ these contributions are logarithmically divergent~\cite{Barbieri:2008cc},  
\beq
\label{e.tparam}
\Delta T = - {3 \over 8 \pi c_W^2} \log (m_\rho/m_Z)  - {3 \over 8 \pi c_W^2} \log (\Lambda/m_\rho)  \left ( 1 - {3 \alpha \over 4} + {\alpha^2 \over 4}  \right ).
\eeq  
The first term in the square bracket is due to  loops with electroweak gauge bosons and the lack of the corresponding Higgs contribution that would cancel it within the SM. 
The second term is due to loops with $\rho$. 
The contributions of \eref{tparam} are always negative, which is disfavored by electroweak precision tests. 
Positive contributions may be obtained by introducing additional degrees of freedom. 
Furthermore, allowing for $\eps \neq 0$ in \eref{opeps} one can obtain quadratically divergent corrections to $T$ which can have either sign. A comprehensive analysis of electroweak precision observables up to one loop  can be found in Ref.~\cite{Cata:2010bv}. 

\section{Broken Parity}\label{sec:broken}

QCD and other vector-like theories at low energies are described by a parity-conserving lagrangian. 
However it is not guaranteed that the dynamics that breaks the electroweak symmetry is vector-like, 
and thus it is conceivable that the interactions of the $\rho$-mesons with the Goldstone bosons do not respect parity (for a concrete example, see  e.g. \cite{Evans:2009ga}).  
In the language of our effective theory, the leading $\co(p^2)$ lagrangian may contain a parity breaking  term,   
\beq
\label{e.LeadingLagrangianPB}
 {v^2 \over 4 (1 -\beta^2)}  \tr \left \{ \alpha V_\mu^+ V_\mu^+  +  V_\mu^- V_\mu^-  - 2 \sqrt{\alpha} \beta V_\mu^- V_\mu^+ 
\right \}  
\eeq 
Here $\beta$ is the order parameter for parity breaking; for $\beta = 0$ we recover the previous parity-conserving case in \eref{LeadingLagrangianPC}.
The positivity of mass and Goldstone kinetic terms requires   $\alpha >0$ and $-1 <\beta < 1 $. 
Furthermore, $\beta \to -\beta$ is a symmetry at $\co(g^0)$, so at that order it is enough to consider $\beta \in (0,1)$. 
The normalization is chosen such that  $v$ is the electroweak scale, that is to say  
$m_{W}^2  =  {g^2 v^2 \over 4} + \co(g^4)$.  
At this point the Goldstone bosons as defined in \eref{xiparam} are not in the right basis:  $\pi^a$ mix with  $\rho^a$  and so they cannot be  interpreted as longitudinal polarizations of $W$ and $Z$ . 
To go to the right basis one needs to make a redefinition, 
\beq
G^a \to  \sqrt{1 - \beta^2}  G^a - \beta  \pi^a   
\eeq  
After this redefinition the Goldstone kinetic terms are canonically normalized, and only $G$ mixes with $\rho$ at the leading order in $g/g_\rho$, 
\beq
\cl \supset  {1 \over 2} (\pa_\mu \pi^a)^2 +  {1 \over 2} (\pa_\mu G^a) ^2 -   m_\rho \rho_\mu^a \pa_\mu G^a  
\eeq 

The most distinctive  phenomenological feature of  the set-up  with a broken parity is the coupling of $\rho$  to 3 pions, 
\beq
{g_{\rho \pi^3} \over 3 v} \left (  \rho_\mu^a \pi^a \pa_\mu \pi^b \pi^b - \rho_\mu^a \pa_\mu \pi^a \pi^b \pi^b \right ) 
\qquad 
g_{\rho \pi^3}  = \beta { \alpha -  \beta^2  \over  \sqrt{\alpha} (1 - \beta^2)} g_{\rho}
\eeq 
This coupling leads to the decay $\rho \to 3 \pi$.  
The widths for the 2- and 3-body decay are given by 
\beq
\Gamma(\rho \to 2 \pi) = {g_{\rho \pi \pi}^2 m_\rho \over 48 \pi}
\qquad 
\Gamma(\rho \to 3 \pi) =  {3 g_{\rho \pi^3}^2 m_\rho^3 \over 4096 \pi^3 v^2}  
\eeq 
The latter is suppressed by 3-body phase space but can be non-negligible in some regions of the parameter space.  

Finally, we discuss the effect of parity violation on the parameter space allowed by unitarity. 
When expressed in terms of $g_\rho$, $g_{\rho \pi \pi}$ and $m_\rho$, the scattering amplitudes for longitudinal $W$, $Z$ and $\rho$  take exactly the same form as in the unbroken parity case, 
 see Eqs. (\ref{e.m0e}),   (\ref{e.m0ie}),  (\ref{e.m0se}). 
However the relation between $m_\rho$ and the couplings is changed for non-zero $\beta$. 
We have, 
\beq
m_{\rho}^2 =   {\alpha g_\rho^2 v^2 \over 1 - \beta^2}  + \co(g^2)
\qquad 
g_{\rho \pi \pi}  =   {\alpha - \beta^2 \over 2 (1 - \beta^2)}  g_\rho
\eeq  
For a given  $g_\rho$ and $g_{\rho \pi \pi}$ the value of $m_\rho$ always increases  compared to the case with $\beta =0$.
This affects the parameter space region allowed by unitarity and typically accelerates unitarity breakdown.  
Furthermore, the processes $\pi\pi \to \pi \rho$ mediated by the $\rho-3\pi$ contact interaction are allowed for $\beta \neq 0$. 
The amplitudes for these processes grow linearly with $s$ above the $m_\rho$ threshold which, for large enough $\beta$, leads to the most stringent 
unitarity bound in certain regions of the parameter space.  
In \fref{unitarity_PB} we plot the contours of the maximum cutoff scale for 2 different values of the parameter $\beta$.  
The theoretically excluded parameter space  where the maximum cutoff is below $m_\rho$ grows  larger as $\beta$ is increased. 
For large enough $\beta$ there is  an upper bound on the coupling $g_\rho$. 
In \fref{unitarity_PB_brs} we see that the branching fraction for the 3-body decay can be up to 30 percent in the parameter space allowed by the unitarity constraints.

\begin{figure}[tb]
\vspace{1cm}
\bc
\includegraphics[width=0.47\textwidth]{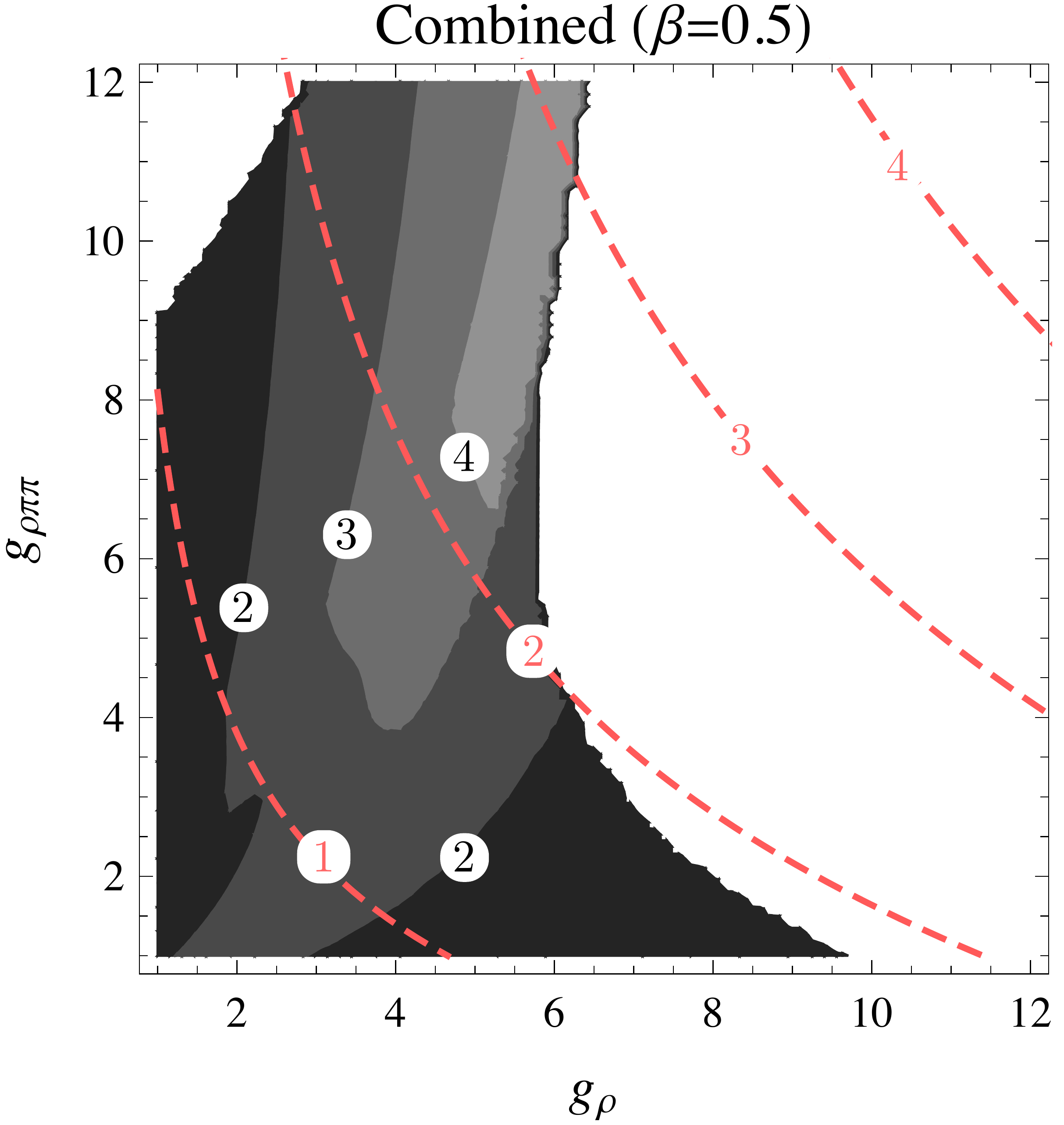}
\hspace{0.5cm} 
\includegraphics[width=0.47\textwidth]{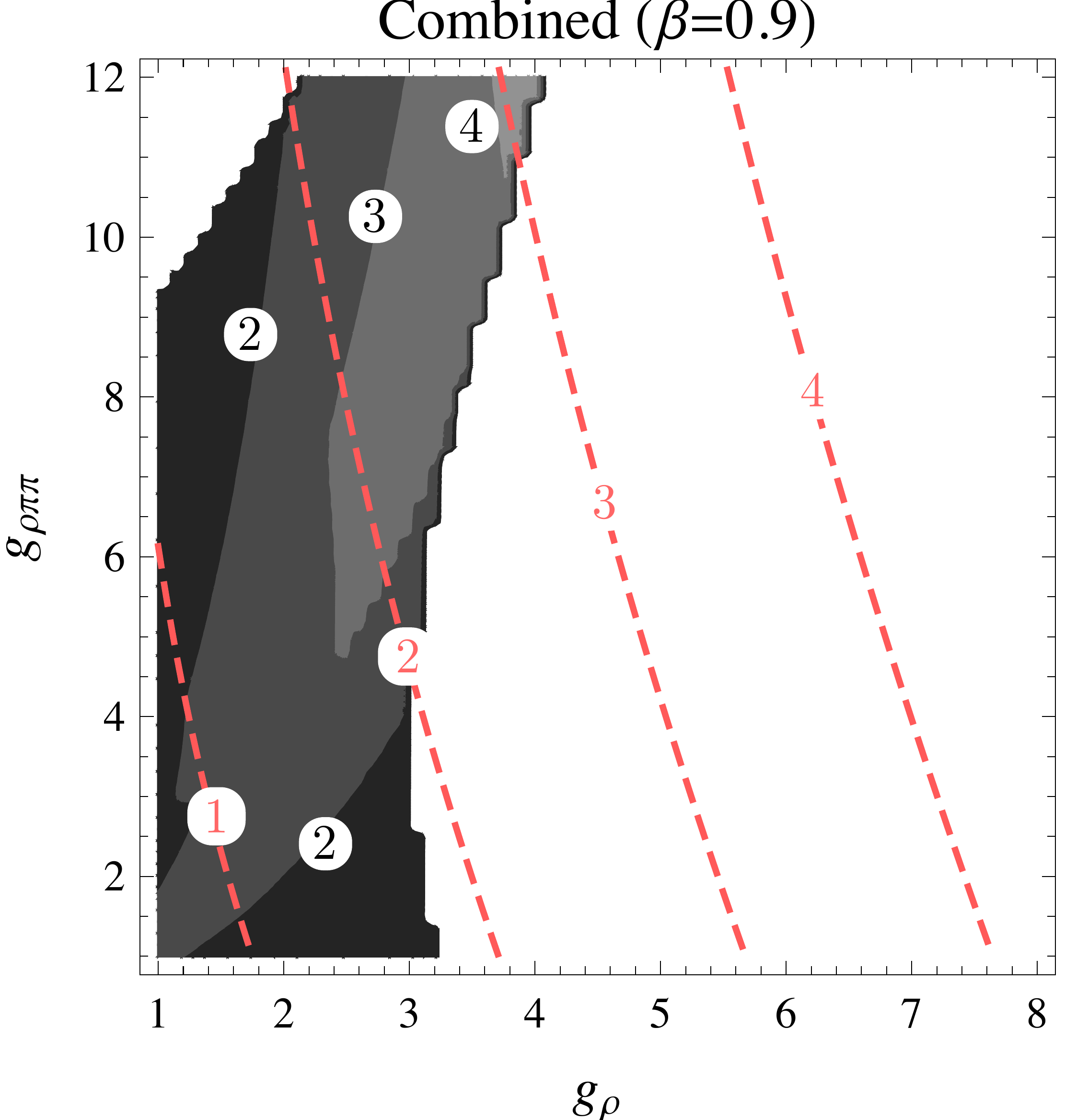}
\ec 
\caption{
\label{f.unitarity_PB}
\small Contour plots of the maximum cut-off scale for $\beta = 0.5$ (left) and $\beta = 0.9$ (right)  
overlaid it with contours of constant $m_\rho$ (dashed). 
The shaded regions correspond to a cutoff scale $\Lambda$ smaller than   2, 3, 4~TeV (from dark to light gray).
}
\end{figure}

The final observation is that parity breaking  affects the tree-level contribution of the $\rho$ mesons to  the $S$ parameter, 
\beq
\Delta S  = {4 \pi \over g_\rho^2} \left (1 - {\beta^2 \over \alpha} \right ) 
\eeq  
For a fixed $g_\rho$, the $S$ parameter is always {\em smaller} than in the $\beta = 0 $ case.  
On the other hand, the allowed parameter space shrinks for large $\beta$, in particular, the region of large $g_\rho$ is not available. 
Nevertheless, in \fref{unitarity_PB_brs} we see that, for moderate $\beta$, $\Delta S \simlt 0.3$ is possible in the allowed parameter space.    

\begin{figure}[tb]
\vspace{0.5cm}
\centerline{
\includegraphics[width=0.47\textwidth]{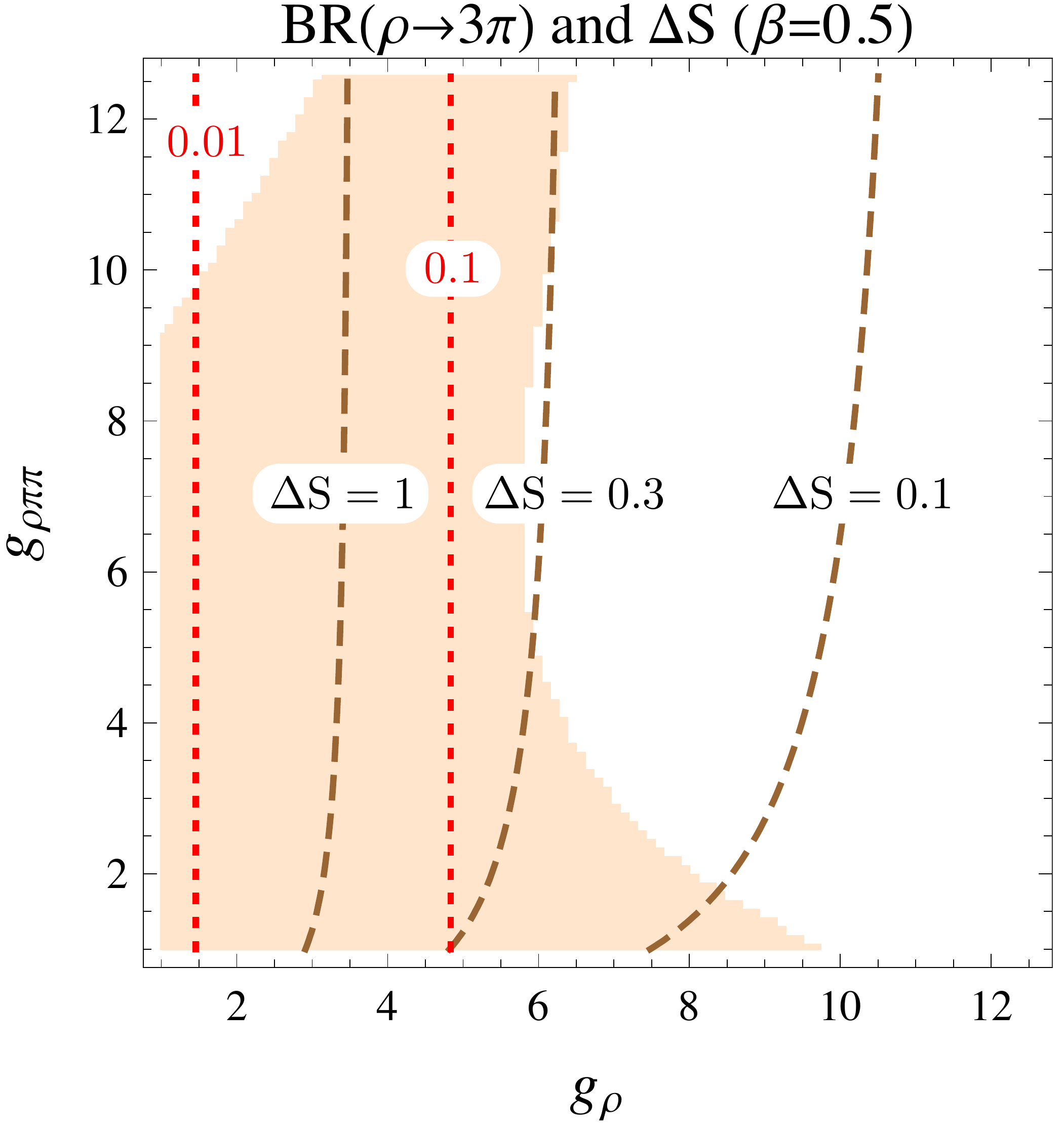}
\hspace{0.5cm} 
\includegraphics[width=0.47\textwidth]{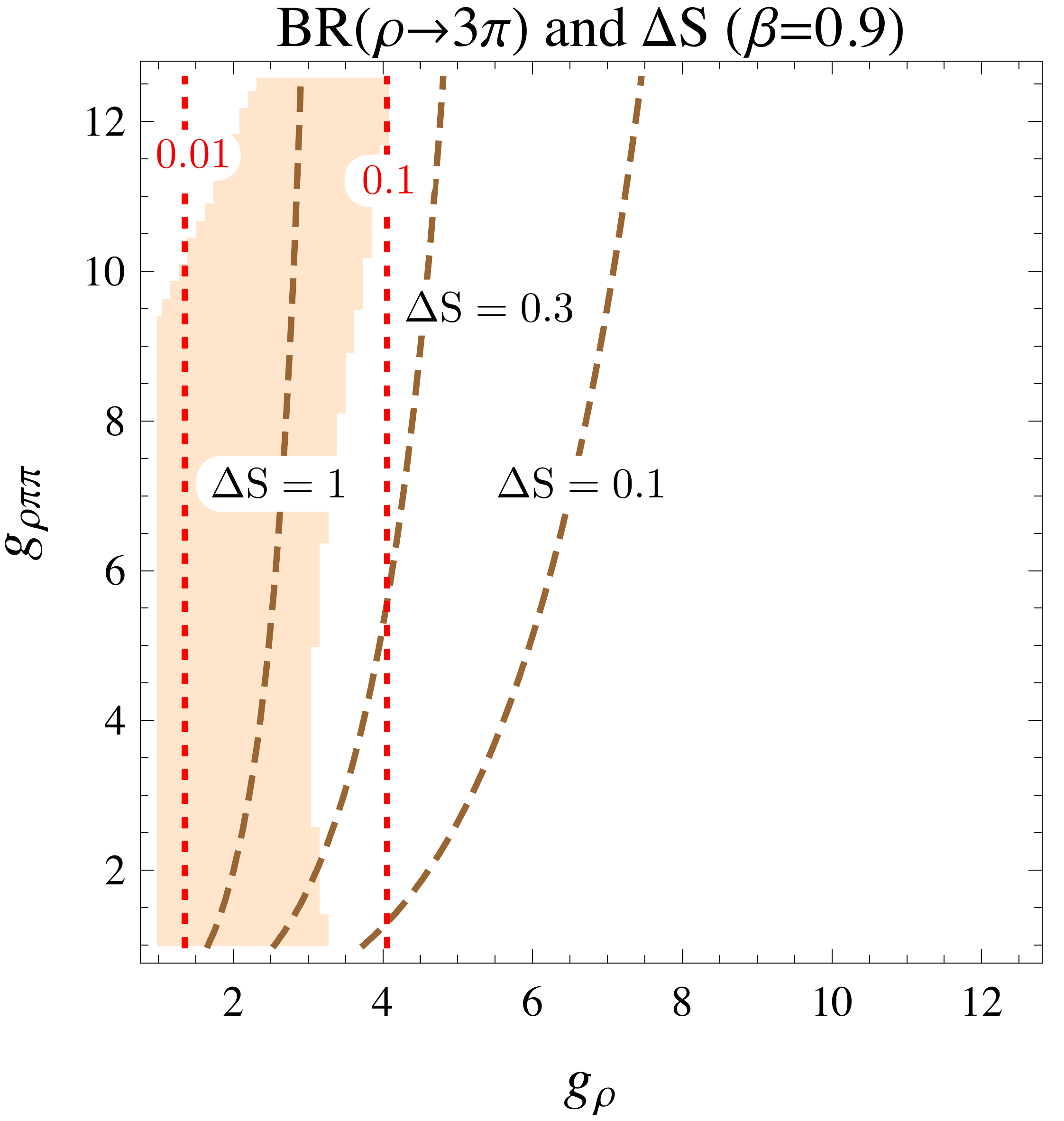}
}
\caption{
\label{f.unitarity_PB_brs}
\small The parameter space allowed by the unitarity constraints (light orange) overlaid with contours of the constant ${\rm Br}(\rho \to 3 \pi)$ (dotted) and
$\Delta S$ (dashed) for $\beta = 0.5$ (left) and $\beta = 0.9$ (right).  
}
\end{figure}

\section{Conclusions}\label{sec:conclusions}

The dynamics of the electroweak symmetry breaking is being probed   at the LHC. A Higgsless scenario, with the electroweak symmetry broken
by a  strongly interacting sector providing the necessary Goldstone bosons to be ``eaten" by the longitudinal $W$ and $Z$, is one of the possibilities.
It can be probed experimentally by discovering the new degrees of freedom linked to the strong dynamics. It is not easy to anticipate the properties
of those degrees of freedom  nor the potential for their discovery at the LHC. We can only be guided by QCD and by various theoretical approaches
to the modelling of nonperturbative effects in strong interactions.  Their main generic common future  is the vector  meson dominance,  i.e., the saturation
of the low energy amplitudes by the lightest resonances that interact perturbatively with the longitudinal $W$ and $Z$.

In this paper we have considered a minimal setup, with a single spin-1 $SU(2)_C$ triplet resonance, $\rho$, in the electroweak symmetry breaking sector,
coupled to the Goldstone bosons (longitudinal $W$ and $Z$) in chiral invariant way. Our goal was to systematically investigate in this framework (and with no
further model dependent assumptions) the parameter space where this theory is under perturbative control and discuss the chances for confirming at the LHC
such a mechanism of electroweak breaking.   Our main conclusions are:
\begin{itemize}

\item The crucial role in determining the range of validity and the cut-off scale is played by  the $\pi\pi \to \rho\rho$ and  $\pi\rho \to \pi\rho$
inelastic channels. A single heavy meson with a mass between 2.5 and 3~TeV  is more efficient in delaying the onset of strong coupling  than a single light resonance with $m_\rho < 2$~TeV.
Still, the maximal value of the cut-off in a single resonance set-up is of the order of $4\pi v$, i.e.,  the NDA result.  

\item Requiring, for consistency of the resonance saturation model, that  the cut-off is  above the resonance
mass and  that the resonance couples perturbatively to $W_LW_L$, the upper bound on the resonance mass is $\mathcal{O}$(3)~TeV.  Thus, if spin-1 resonances play the dominant role in pushing  the perturbative unitarity bound in the $W_LW_L$ scattering beyond 1.7~TeV (the perturbative unitarity cut-off in the ``Higgless" SM), a resonance must exist with mass below $\mathcal{O}$(3)~TeV.  If a lighter resonance is found, the perturbative unitarity cut-off obtained with such a single resonance is lower than the discussed above maximal cut-off
but may be well above the resonance mass (see Fig.~5).  This opens up the possibility of more resonances playing a role in perturbative unitarization of the $W_LW_L$ scattering.

\item An interesting parameter is the triple-$\rho$ coupling $g_\rho$ and its correlation with $m_\rho$, expressed by the equation $m^2_\rho=\alpha g^2_\rho v^2$. The QCD value $\alpha\sim 2$
is not the one that maximizes  the cut-off for a given value of the resonance mass.

\item Relaxing the hypothesis of parity invariance in the strong sector allows for a small $S$ parameter but does not increase the region of perturbative control of the model.

\item The LHC in its high-energy phase will explore a large fraction of the parameter space of the spin-1 resonances compatible with perturbative unitarity. Nonetheless, if the resonances are heavy ($m_\rho \simgt 2$ TeV) and strongly coupled ($g_\rho \simgt 6$) the searches become very challenging and they might escape any direct detection. 

\end{itemize}

In the presence or the absence of a light Higgs boson, the measurement of the $W_LW_L$ scattering amplitude is of prime importance to decipher the true dynamics of the electroweak symmetry breaking. Below the resonance masses, the amplitudes we studied can be casted in the form of a chiral expansion with $p^4$ contact interactions among the pions: $\alpha_4=-\alpha_5=\frac{g_{\rho\pi\pi}^2}{4}\frac{v^4}{m_\rho^4}$. It should be noted however  that the standard procedures to unitarize such a model, such as Pad\'e unitarization scheme,  do not reproduce the parameters of the resonance amplitude we started with and can lead to unphysical results and interpretations. Therefore the simple formalism to describe spin-1 resonances developed in this paper might be crucial in determining the true agents responsible for the breaking of the electroweak symmetry.

\appendix

\section{Identifying physical degrees of freedom}\label{sec:eigenstates}

In this appendix we identify the combinations of fields  in the ``gauge" basis  that correspond to vector boson mass eigenstates and their Goldstone bosons. 

 \bc \bf Vector boson mass eigenstates  \ec

We start with the lagrangian \eref{LeadingLagrangianPC}. 
The mass terms for the charged vector bosons  are given by   
\beq
\cl_{mass} = {g^2 v^2 \over 8}  L_\mu^i L_\mu^i  +  {\alpha  v^2 \over 2 }   [2  g_\rho \rho_\mu^i -  g L_\mu^i]^2 \, . 
\eeq   
Due to the second term the mass matrix is non-diagonal in the gauge basis. 
To go to the mass eigenstate basis we need the rotation  
\beq
\label{e.chargedrotation} 
\bvec L^i  \\ \rho^i \evec \to 
R_c \cdot \bvec W^i  \\ \rho_c^i \evec 
\qquad 
R_c = \left (\ba{cc} \cos x_c & -\sin x_c  \\ \sin x_c & \cos x_c \ea \right )    
\eeq 
with the rotation angle given by  
\beq 
\tan 2 x_c =  {g \over g_\rho} \left ( 1   -  {1 + \alpha \over \alpha}  {g^2 \over 4 g_\rho^2} \right )^{-1}   \, . 
\eeq 
We consider  the parameter range with $\alpha \sim \co(1)$ and $g \ll g_\rho$ in which case we can expand in powers  of $g/g_\rho$. 
The mass eigenvalues and the rotation angle can be approximated by
\bea
\label{e.Wmixing}
m_{\rho_c}^2 & \approx & \alpha g_\rho^2 v^2 \left (1 + {g^2 \over 4 g_\rho^2}  \right ) ,
\nn 
m_W^2 & \approx & {g^2 v^2 \over 4} \left (1 -  {g^2 \over 4 g_\rho^2}  \right),
\nn 
\sin x_c  &\approx&  {g \over 2 g_\rho} \left (1   -  {g^2 \over 4 g_\rho^2}   {\alpha - 2 \over 2 \alpha } \right ) .
\eea 
When $g \ll g_\rho$ the eigenvalues are hierarchical.  Moreover, the gauge basis is related to the mass eigenstate basis by a parametrically small rotation  $\sim g/2 g_\rho$. 
  
For the neutral vector bosons the mass terms are 
\beq
\cl_{mass} = {v^2 \over 8} \left ( [g L_\mu^3 - g' B_\mu]^2 + \alpha [2  g_\rho \rho_\mu^3 -  g L_\mu^3 - g' B_\mu]^2  \right ).
\eeq   
These mass terms yield a massless eigenstate corresponding to the ordinary photon. 
To isolate it,  one starts with the usual SM rotation $L^3 \to (g' \ti A + g \ti Z  )/\sqrt{g^2+g'^2}$,  $B \to (g \ti A - g' \ti Z  )/\sqrt{g^2+g'^2}$. 
However, $\ti A$ is not the physical photon yet because it mixes with $\rho^3$. 
The massless photon  is defined by the rotation 
$\rho^3  \to  (g_\rho \ti \rho + e  A)/\sqrt{g_\rho^2+e^2}$,
$\ti A  \to  (-e \bar \rho + g_\rho   A)/\sqrt{g_\rho^2+e^2}$, where $e = g g'/\sqrt{g^2+ g'{}^2}$.  
Finally, we need a 2D rotation in the plane $\ti \rho, \ti Z$  to arrive at the mass eigenstates.   
Summarizing these 3 rotations,  
\beq
\bvec L^3 \\ B \\ \rho^3 \evec \to 
\left ( \ba{c|c|c}
{g  \cos x_n \over \sqrt{g^2 + g'{}^2}} - {g' \sin x_n \over \sqrt{ g^2 + g'{}^2}} {e \over \sqrt{g_\rho^2 + e^2}} 
& {g' \over \sqrt{g^2 + g'{}^2}} {g_\rho \over \sqrt{g_\rho^2 + e^2}}
& - {g  \sin x_n \over \sqrt{g^2 + g'{}^2}}- {g' \over \sqrt{g^2 + g'{}^2}} {e \cos x_n  \over \sqrt{g_\rho^2 + e^2}} 
\\
\hline 
- {g'  \cos x_n \over \sqrt{g^2 + g'{}^2}} - {g \sin x_n \over \sqrt{ g^2 + g'{}^2}} {e \over \sqrt{g_\rho^2 + e^2}} 
& {g \over \sqrt{g^2 + g'{}^2}} {g_\rho \over \sqrt{g_\rho^2 + e^2}}
&  {g'  \sin x_n \over \sqrt{g^2 + g'{}^2}}- {g \over \sqrt{g^2 + g'{}^2}} {e \cos x_n  \over \sqrt{g_\rho^2 + e^2}} 
\\
\hline 
{g_\rho \over \sqrt{g_\rho^2 + e^2}} \sin x_n
&  {e \over \sqrt{g_\rho^2 + e^2}}
& {g_\rho \over \sqrt{g_\rho^2 + e^2}} \cos x_n
\ea \right ) 
\bvec  Z \\  A \\ \rho_0 \evec  .
\eeq 
Expanding the eigenvalues and the rotation angle in  $g/g_\rho$, 
\bea
\label{e.Zmixing}
m_{\rho 0}^2 & \approx & \alpha g_\rho^2 v^2 \left ( 1 + {g^2 +g'{}^2\over 4 g_\rho^2}  \right ), 
\nn 
m_Z^2 & \approx & {g^2  +g'{}^2 \over 4} v^2 \left (1 -  {(g^2- g'{}^2)^2 \over 4 (g^2 + g'{}^2) g_\rho^2}  \right),
\nn 
\sin x_n &\approx&  { g^2- g'{}^2 \over 2 \sqrt{ g^2 +g'{}^2}g_\rho} \left (1   -  {g^2 +g'{}^2 \over 4 g_\rho^2}   {\alpha - 2 \over 2 \alpha } \right ).
\eea   

\vspace{.2cm}
\bc \bf Goldstone boson eigenstates \ec 
 
The original lagrangian  \eref{LeadingLagrangianPC} leads to diagonal kinetic terms and no mass terms for the Goldstone bosons $\pi$ and $G$. 
However, mass mixing between $\pi$ and $G$ appears after adding the gauge fixing term\footnote{Equivalently, one can define Goldstone boson eigenstates as the linear combinations  that diagonalize  the kinetic mixing with the vector eigenstates. That procedure would lead to the same expression for Goldstone boson eigenstates. } 
As is customary, we choose the $R_\xi$ gauge fixing terms such that the kinetic mixing between the gauge and Goldstone fields is removed, 
\beq
\label{e.gaugefixing}
  \cl_{gf} = {1 \over 2 \xi}  \left[ \pa_\mu L_\mu^a -   \xi {g v \over 2} (\pi^a - \sqrt \alpha G^a) \right]^2 
+ {1 \over 2 \xi}  \left[ \pa_\mu B_\mu -   \xi {g' v \over 2} (\pi^3 +  \sqrt \alpha G^3) \right]^2 
+ {1 \over 2 \xi} \left[ \pa_\mu \rho_\mu^a -  \xi \sqrt{\alpha} g_\rho v G^a \right]^2  . 
\eeq 
Now the Goldstone mass terms are not diagonal. 
To arrive at the  mass eigenstate basis we need the rotation  
\bea 
\pi^ i\to \cos y_c \pi_c^i - \sin y_c  G_c^i  &\qquad&  \pi^ 3 \to \cos y_n \pi_0 - \sin y_n  G_0 ,
\nn 
G^i \to  \sin y_c  \pi_c^i + \cos y_c  G_c^i &\qquad&   G^3 \to  \sin y_n \pi_0 + \cos y_n G_0  ,
\eea 
\bea 
\tan 2 y_c &=&  {g^2 \over 2 \sqrt \alpha g_\rho^2} \left ( 1 + {\alpha - 1 \over \alpha} {g^2 \over 4 g_\rho^2} \right )^{-1},
\nn 
\tan 2 y_n &=&  {g^2  - g'{}^2 \over 2 \sqrt \alpha g_\rho^2} \left ( 1 + {\alpha - 1 \over \alpha} {g^2  + g'{}^2 \over 4 g_\rho^2} \right )^{-1}. 
\eea 
After these rotations one finds that the Goldstone boson mass eigenvalues  are  {\em exactly} $\xi^{1/2}$ times the corresponding vector masses: 
$m_{\pi_c}^2 = \xi m_W^2$, $m_{\pi_0}^2 = \xi m_Z^2$, $m_{G_c}^2 = \xi m_{\rho_c}^2$,  $m_{G_0}^2 = \xi m_{\rho_0}^2$.  
The original Goldstone boson basis  is related  to the mass eigenstates basis via a rotation with the angle suppressed by $g^2/g_\rho^2$. 
In the main body of the paper we employed the Goldstone  bosons to compute the scattering amplitudes of  longitudinally polarized vector bosons. 
These amplitudes start at $(g/g_\rho)^0$, and at that order it is sufficient to use Goldstone and vector fields in the original gauge basis.

\section{Deconstructed (gauge) models}\label{sec:deconstruction}

\begin{figure}[tb]
\vspace{1cm}
\bc
\includegraphics[width=0.5\textwidth]{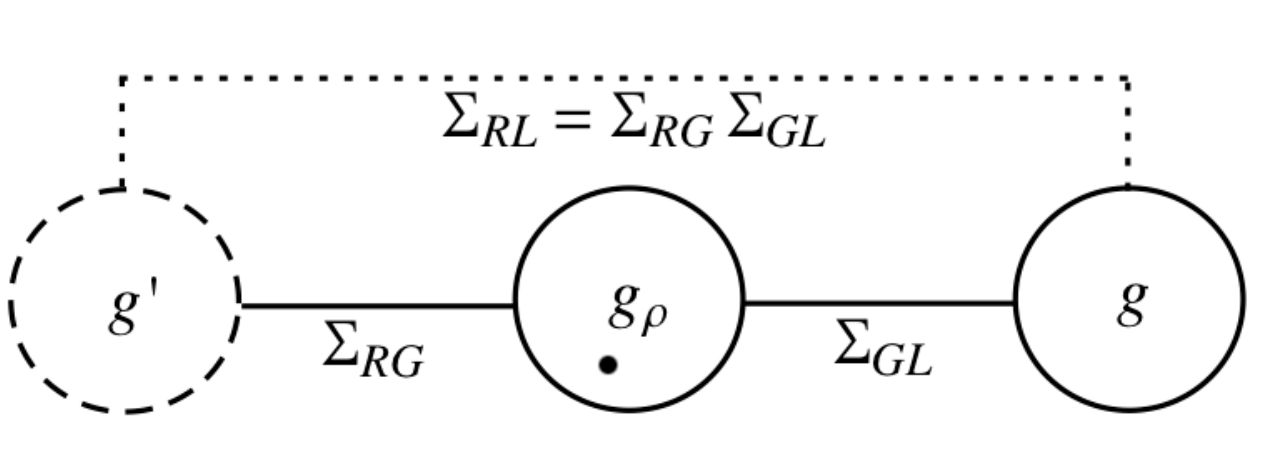}
\ec 
\caption{ 
\label{f.moose}
\small Moose diagram for the most general three-site model. 
}
\end{figure}

The most general single resonance ``gauge" model with global $G=SU(2)_{R}\times SU(2)_{G}\times SU(2)_{L}$ spontaneously broken to the custodial subgroup $H=SU(2)_{C}$ and with the $SU(2)_{G}$ subgroup fully gauged can be described in terms of the following $\sigma$-model Lagrangian
\bea
\mathcal{L}=\hat{v}^{2}\left< D^{\mu}\Sigma_{RG} D_{\mu}\Sigma_{RG}^{\dag} \right>+\hat{v}^{2}\left( 1+\hat{\epsilon}\right) \left< D^{\mu}\Sigma_{GL} D_{\mu}\Sigma_{GL}^{\dag} \right>
\nn
+\hat{\delta}\hat{v}^{2}\left\langle \Sigma_{GL}\left( D_{\mu}\Sigma_{GL}^{\dag}\right) \Sigma_{RG}^{\dag}\left( D^{\mu}\Sigma_{RG}\right) \right\rangle +\mathcal{L}_{gauge\; kinetic}.
\label{lagparz}
\eea
The structure of the model can be illustrated by a moose diagram presented in Fig.~\ref{f.moose}.
The $\Sigma_{ij}$ fields transform as $\Sigma_{ij}\ \rightarrow\ g_{i}\Sigma_{ij} g_{j}^{\dag}$ where $g_{i,\;j}$ are elements of the various $SU(2)$. The covariant derivatives are given by
\bea
D_{\mu}\Sigma_{RG}=\partial_{\mu}\Sigma_{RG}-i {g' \over 2}  B_\mu  \sigma^3 \Sigma_{RG}+i \Sigma_{RG} {g_\rho \over 2}  \rho_\mu^a \sigma^a,
\nn
D_{\mu}\Sigma_{GL}=\partial_{\mu}\Sigma_{GL}-i {g_\rho \over 2}  \rho_\mu^a \sigma^a  \Sigma_{GL}+i\Sigma_{GL} {g \over 2}  W_\mu^a \sigma^a.
\eea
The non-standard term $\left\langle \Sigma_{GL}\left( D_{\mu}\Sigma_{GL}^{\dag}\right) \Sigma_{RG}^{\dag}\left( D^{\mu}\Sigma_{RG}\right) \right\rangle $ introduces a non-local interaction and leads to the most general form of the three-site model.  In the above model parity is not assumed - parity violation is described by $\hat{\epsilon}$. It might be also useful to note that $\Sigma_{RG}\Sigma_{GL}=U$, which is an object often used in chiral perturbation theory.

One can make an immediate connection with the ``hidden gauge" formalism by observing that 
\begin{equation}
\xi_{R}=\Sigma_{RG},\ \ \ \ \ \ \ \ \xi_{L}=\Sigma_{LG}=\Sigma_{GL}^{\dag}
\end{equation}
and
\bea
\left\langle D^{\mu}\Sigma_{RG}D_{\mu}\Sigma^{\dag}_{RG}\right\rangle =-\frac{1}{2}\left\langle V^{-}_{\mu}V^{+}_{\mu}\right\rangle +\frac{1}{4}\left\langle \left( V^{+}_{\mu}\right)^{2} \right\rangle +\frac{1}{4}\left\langle \left( V^{-}_{\mu}\right)^{2} \right\rangle ,
\nn
\left\langle D^{\mu}\Sigma_{GL}D_{\mu}\Sigma^{\dag}_{GL}\right\rangle =\frac{1}{2}\left\langle V^{-}_{\mu}V^{+}_{\mu}\right\rangle +\frac{1}{4}\left\langle \left( V^{+}_{\mu}\right)^{2} \right\rangle +\frac{1}{4}\left\langle \left( V^{-}_{\mu}\right)^{2} \right\rangle ,
\nn
\left\langle \Sigma_{GL}\left( D_{\mu}\Sigma_{GL}^{\dag}\right) \Sigma_{RG}^{\dag}\left( D^{\mu}\Sigma_{RG}\right) \right\rangle =\frac{1}{4}\left\langle \left( V^{-}_{\mu}\right)^{2} \right\rangle -\frac{1}{4}\left\langle \left( V^{+}_{\mu}\right)^{2} \right\rangle .
\eea
Then the ``gauge" model Lagrangian (\ref{lagparz}) is equivalent to the general ``hidden gauge" Lagrangian (\ref{e.LeadingLagrangianPB}) with
\begin{equation}
v^{2}=4 \hat{v}^{2}\left( \frac{\hat{\delta}}{2}+\frac{\left( 1+\frac{\hat{\delta}}{2}\right) \left( 1+\hat{\epsilon}+\frac{\hat{\delta}}{2}\right) }{2+\hat{\delta}+\hat{\epsilon}}\right) ,\ \ \ \alpha=\frac{2+\hat{\delta}+\hat{\epsilon}}{2+3\hat{\delta}+\hat{\epsilon}},\ \ \ \beta=\frac{-\hat{\epsilon}}{\sqrt{\left( 2+3\hat{\delta}+\hat{\epsilon}\right) \left( 2+\hat{\delta}+\hat{\epsilon}\right) }}.
\end{equation}
resulting in the same $g_{\rho\pi\pi}$ coupling and the same form of elastic and inelastic $WW$ scattering amplitudes.
The local three-site model~\cite{Chivukula:2006cg} clearly predicts $\alpha=1$.

\section*{Acknowledgments} 

We thank Tulika Bose, Michele Papucci, Slava Rychkov, and Jure Zupan for useful discussions. 
This work has been partially  supported by the European Commission under the contract ERC Advanced Grant 226371 MassTeV, the contract PITN-GA-2009-237920 UNILHC and the MNiSzW scientific research grant 
N202 103838 (2010 - 2012). The work of A.W. was supported in part by the German Science Foundation (DFG) under the Collaborative Research Center (SFB) 676.


\end{document}